\documentclass[10pt,twocolumn,letterpaper]{article}

\usepackage[pagenumbers]{cvpr} % 

\usepackage{amsmath}
\usepackage{amsthm}

\usepackage{algorithm}
\usepackage{algpseudocode} % From the algorithmicx package
\usepackage{booktabs} % 引入 booktabs 宏包以使用 \toprule, \midrule, \bottomrule
\usepackage{tabularx}

\usepackage[a4paper, margin=1in]{geometry}
\usepackage{graphicx}
\usepackage[table]{xcolor}
\usepackage{color}

\newtheorem{theorem}{Theorem}
\newtheorem{definition}[theorem]{Definition}
\newtheorem{lemma}[theorem]{Lemma} % Defines the 'lemma' environment
\newtheorem{proposition}[theorem]{Proposition}
\newtheorem{remark}[theorem]{Remark}
\newtheorem{corollary}[theorem]{Corollary}

\newcommand{\appdx}[1]{{\color{black}}}

\definecolor{cvprblue}{rgb}{0.21,0.49,0.74}
\usepackage[pagebackref,breaklinks,colorlinks,allcolors=cvprblue]{hyperref}
\usepackage[T1]{fontenc}    % use 8-bit T1 fonts
\usepackage{textcomp}
\usepackage{gensymb}
\usepackage{makecell,multirow,diagbox,colortbl}
\usepackage{amsthm} % new add

\usepackage{balance} % new add

\title{One Walk is All You Need: Data-Efficient 3D RF Scene Reconstruction \\ with Human Movements
}

\author{
    Yiheng Bian\textsuperscript{1} \quad 
    Zechen Li\textsuperscript{1} \quad   
    Lanqing Yang\textsuperscript{1}\thanks{Corresponding author.} \quad 
    Hao Pan\textsuperscript{1} \quad 
    Yezhou Wang\textsuperscript{1} \quad 
    Longyuan Ge\textsuperscript{1} \quad 
    \\ Jeffery Wu\textsuperscript{1} \quad Ruiheng Liu\textsuperscript{1} \quad Yongjian Fu\textsuperscript{2} \quad Yichao Chen\textsuperscript{1} \quad Guangtao Xue\textsuperscript{1} \\[1ex]
    \textsuperscript{1}Shanghai Jiao Tong University \quad \textsuperscript{2}Central South University\\
    {\tt\small \{byhbye123, yanglanqing, zechlee, panhao09, yezhouwang, gly2000,}\\
    {\tt\small jeffery2019, liutuiheng, yichao, gt\_xue\}@sjtu.edu.cn, fuyongjian@csu.edu.cn}
}

\begin{document}
\maketitle

\begin{abstract}

Reconstructing 3D Radiance Field (RF)  scenes through opaque obstacles is a long-standing goal, yet it is fundamentally constrained by a laborious data acquisition process requiring thousands of static measurements, which treats human motion as noise to be filtered.
This work introduces a new paradigm with a core objective: to perform fast, data-efficient, and high-fidelity RF reconstruction of occluded 3D static scenes, using only a single, brief human walk. 
We argue that this unstructured motion is \textbf{not noise}, but is in fact an \textbf{information-rich signal} available for reconstruction. 
To achieve this, we design a factorization framework based on composite 3D Gaussian Splatting (3DGS) that learns to model the dynamic effects of human motion from the persistent static scene geometry within a raw RF stream.
Trained on just a single 60-second casual walk, our model reconstructs the full static scene with a Structural Similarity Index (SSIM) of \textbf{0.96}, remarkably outperforming heavily-sampled state-of-the-art (SOTA) by \textbf{12\%}. 
By transforming the human movements into its valuable signals, our method eliminates the data acquisition bottleneck and paves the way for on-the-fly 3D RF mapping of unseen environments.

\end{abstract}
\section{Introduction}
\label{sec:introduction}

The unique ability of Radiance Field (RF) signals to penetrate opaque obstacles offers a new frontier for building high-fidelity 3D models of non-line-of-sight (NLOS) spaces, with critical applications in robotics, augmented reality, and smart infrastructure~\citep{adib2013see, vasisht2016decimeter}.

However, this promising frontier has been fundamentally constrained by a bottleneck: a costly, brute-force approach to data acquisition. Whether using early tomographic techniques~\citep{adib2013see} or modern neural representations like Neural Radiance Fields (NeRF)~\citep{mildenhall2021nerf, lu2024newrf} and 3D Gaussian Splatting (3DGS)~\citep{wen2024wrf, zhang2024rf}, all existing methods are slaves to a cripplingly laborious process. They rely on robotic platforms to scan dense grids over hours, capturing thousands of calibrated static measurements to disambiguate the scene's geometry. This insatiable demand for data has relegated high-fidelity RF mapping to the laboratory, making rapid, real-world deployment an impossibility.

The field's foundation rests on a seemingly indisputable dogma: the reconstruction of static scenes must rely on \textbf{static measurements}, while \textbf{dynamic events}, especially unpredictable human motion, are a source of chaotic noise that must be aggressively filtered or entirely avoided~\citep{qian2018enabling}.

In this paper, we fundamentally challenge this long-standing dogma and posit the contrary: the chaotic interference from unstructured human motion is not a source of noise to be mitigated, but rather an information-rich signal for reconstructing the static world.
A person walking through a room is not an obstacle to be mitigated, but a powerful, active probe. Each step casts a unique ``RF shadow'', creating a cascade of complex diffractions and occlusions. This single, continuous dynamic event implicitly scans the environment from \textbf{thousands of virtual viewpoints}, revealing geometric details of the static background that a sparse grid of static measurements could never see. \textbf{The ``noise'' is the solution.}

Our argument begins with a new theoretical foundation. We provide an analysis demonstrating why a single, casual one-minute walk can provide geometric information equivalent to deploying a dense array of up to 10x more physical RF sensors. Critically, we then demonstrate that this rich, dynamic information can only be effectively harnessed by models exhibiting \textbf{linear superposition}. The complex, additive nature of multi-path scattering from a static background and a dynamic human necessitates a model that can linearly combine these components without destructive interference.

This principle leads us directly to 3D Gaussian Splatting (3DGS), whose rendering process, a linear summation of Gaussian contributions, is ideally suited for this task. Based on this insight, we propose a principled disentanglement framework. Instead of training a single monolithic model, we employ a \textbf{two-stage strategy}. First, we train a dense background 3DGS model on a minimal static dataset to represent the background environment. Second, we introduce a new, sparse set of ``human Gaussians'' and train them exclusively on the dynamic RF stream together with frozen background model , allowing them to capture the human-induced perturbations. The final, high-fidelity scene is rendered by simply adding these two linear models together, a process that inherently preserves the integrity of the static background while incorporating the rich geometric cues from the human motion.

Experiments on $3$ different scenes show that our method significantly outperforms state-of-the-art (SOTA) baselines. More tellingly, we show that all SOTA methods improve when trained on the human-present data using our framework, proving the intrinsic value of the motion-induced signal. Our extensive ablations reveal crucial insights:
(i) The performance gain does not stem from merely increased data volume, but from the \textbf{structured information within the perturbations}.
(ii) Naive \textbf{end-to-end approach fails}, as it catastrophically overfits to transient noise; our two-stage framework is essential to isolate the signal.
(iii) The primary \textbf{benefit arises from rich, diffuse scattering}, not simple specular reflections, and is most effective when the person's path intersects the transceivers in compact spaces.

Our contributions can be summarized as follows:
\begin{itemize}
\item We are the first to propose a new paradigm for RF reconstruction that repurposes unstructured human motion from debilitating noise into the primary signal for modeling the static environment.
% \item We theoretically and empirically demonstrat that human motion is the optimal signal source, not noise.
\item We propose a principled framework based on the linear superposition property of 3DGS, enabling the effective disentanglement of static and dynamic scene components.
\item Experiments show that we achieve Structure Similarity Index Measure (SSIM) of up to $0.96$,  surpassing heavily-sampled static SOTA methods by $12\%$, showing that a single 60-second walk can replace massive sensor arrays, thereby eliminating the data acquisition bottleneck. 
\end{itemize}

\section{Related Work}

\textbf{Traditional RF Scene Reconstruction.} Modeling the physical world with RF signals is a fundamentally challenging, ill-posed problem due to the complex nature of multi-path propagation~\citep{jiang2023fisherrf,jiang2024terahertz, wei2023integrated, banerjee2024horcrux}. Consequently, the dominant paradigm for high-fidelity reconstruction has been to overcome this ambiguity through dense spatial sampling. Early works in computational imaging used techniques akin to tomography, requiring extensive measurements to form a coherent image~\citep{adib2013see}. Even modern methods, which aim to reconstruct detailed channel information like the power angular spectrum (PAS), are bound by this constraint~\citep{kouyoumjian2005uniform,na2022huygens,maxwell1873treatise}. They typically rely on robotic platforms to meticulously scan an environment, capturing thousands of data points from a dense grid of known locations to solve the complex inverse problem. %This laborious data acquisition process remains the single greatest bottleneck preventing the widespread, practical application of RF-based 3D reconstruction.

\textbf{Neural Representations for RF Scene Modeling.} The rise of neural fields has offered a powerful new toolset for this task. Inspired by the success of Neural Radiance Fields (NeRF)\citep{mildenhall2021nerf} in computer vision, researchers have adapted these techniques to the RF domain. Works like NeRF2\citep{zhao2023nerf2}, NeWRF~\citep{lu2024newrf}, and WiNeRT~\citep{orekondy2023winert} have demonstrated the ability to learn a continuous representation of an RF scene from discrete samples, enabling interpolation of channel properties at unmeasured locations. More recently, following the trend in vision, methods have shifted towards 3D Gaussian Splatting (3DGS)\citep{kerbl20233d,fan2024lightgaussian,fang2024mini,girish2024eagles,lee2024compact3d,niemeyer2024radsplat}for its superior training speed and rendering efficiency. WRF-GS\citep{wen2024wrf} and RF-3DGS~\citep{zhang2024rf} have successfully replaced the MLP backbone of RF NeRFs with 3D Gaussians, achieving comparable or better quality with significantly reduced computational cost. However, a critical thread unites all these advanced methods: they are designed exclusively for static scenes and are predicated on the availability of the same dense, static, and painstakingly collected datasets. %They inherit, rather than solve, the data acquisition bottleneck.

\textbf{Handling Dynamics in Scene Representation.} The concept of dynamics has been treated in two starkly different ways by the vision and RF communities. In computer vision, dynamic scenes are a primary object of study. Methods like D-NeRF~\citep{pumarola2021dnerf} and Dynamic 3D Gaussians~\citep{luiten2023dynamic} explicitly model the motion of objects and people with the goal of reconstructing and rendering the dynamic elements themselves. \appdx{The motion is the signal of interest. In stark contrast, the RF sensing community has historically treated dynamic events, especially human motion, as a source of noise to be either filtered or exploited for a different task. For instance, } Researchers have worked to isolate the minute signal variations from breathing and heartbeats by aggressively filtering out the larger interference from body motion~\citep{qian2018enabling}, or have used the macro-level disturbances to classify activities or detect presence~\citep{ding2024milliflow}. In all cases, the dynamic interference is either a nuisance to be removed or a low-resolution signal for a classification task, not a tool for high-fidelity reconstruction of the surrounding static world.

\appdx{For comparison, this work resides at the intersection of these fields but charts a completely new course. We are the first to propose that the unstructured, dynamic interference from human motion should not be filtered, avoided, or used for classification, but rather harnessed as the primary signal for reconstructing the static scene. Unlike dynamic scene modeling in vision, we do not aim to render the moving person; we treat them as a latent variable. Unlike prior RF sensing work, we do not treat their motion as noise; we treat it as a rich, structured probe. Our unsupervised factorization framework is the first to learn this disentanglement, using the  ``noise'' to solve the data acquisition bottleneck.}
\section{Background}
\label{sec:physical_principle}
% \section{RF Propagation Formulas}

\subsection{Primer on Electromagnetics}
% \label{sec:physical_principle}

Radiance field reconstruction requires modeling complex propagation behaviors such as reflection, transmission, refraction, diffraction, and absorption, alongside multipath effects. We quantify these behaviors using the following mathematical models.

\subsection{Propagation Formulations}

\textbf{Path Loss.}
The attenuation of EM waves over a distance $d$ is calculated as:
    $\text{pathloss} = 20 \log_{10} \frac{4\pi d f}{c}$,
where $c$ is the speed of light and $f$ is the frequency.

\textbf{Reflection, Transmission, and Absorption.}
When waves encounter an obstacle, their behavior is governed by the material's permittivity ($\epsilon_m$) and conductivity ($\mu_m$). Using equivalent circuit models\cite{Hu2012LiIon}, we derive the reflection ($R$), transmission ($T$), and absorption ($A$) rates:
\begin{align}
    R &= \left| \frac{\sqrt{\eta_m}-\sqrt{\eta_0}}{\sqrt{\eta_m}+\sqrt{\eta_0}} \right|^2 \\
    T &= \left|\frac{2\sqrt{\eta_0}}{\sqrt{\eta_m}+\sqrt{\eta_0}}\right|^2 \\
    A &= 1 - R - T
\end{align}
where $\eta_m = \frac{\mu_m}{\epsilon_m}$ and $\eta_0 = \frac{\mu_0}{\epsilon_0}$ are the impedances of the material and air. 

The relationship between the angle of refraction ($\theta_t$) and incidence ($\theta_i$) is:
\vspace{-0.1in}
\begin{equation}
    \frac{\sin(\theta_t)}{\sin(\theta_i)} = \sqrt{\frac{\epsilon_m \mu_m}{\epsilon_0 \mu_0}}
\end{equation}
\vspace{-0.1in}

\textbf{Diffraction.}
Edge diffraction is modeled using the Uniform Theory of Diffraction (UTD)\citep{tsingos2001modeling}. The diffracted field $E_d$ relates to the incident field $E_i$ via:
\begin{equation}
    E_d = E_i D(\theta_i, \theta_d, k, \epsilon_m, \mu_m) F
\end{equation}
Here, $D(\cdot)$ is the diffraction coefficient, $F$ is a polarization factor, and $k = \frac{2\pi f}{c}$ is the wavenumber.

\subsection{Power Angular Spectrum (PAS)}

\begin{figure}[t]
    \setlength{\abovecaptionskip}{0pt}
  \centering
  \includegraphics[width=1\linewidth]{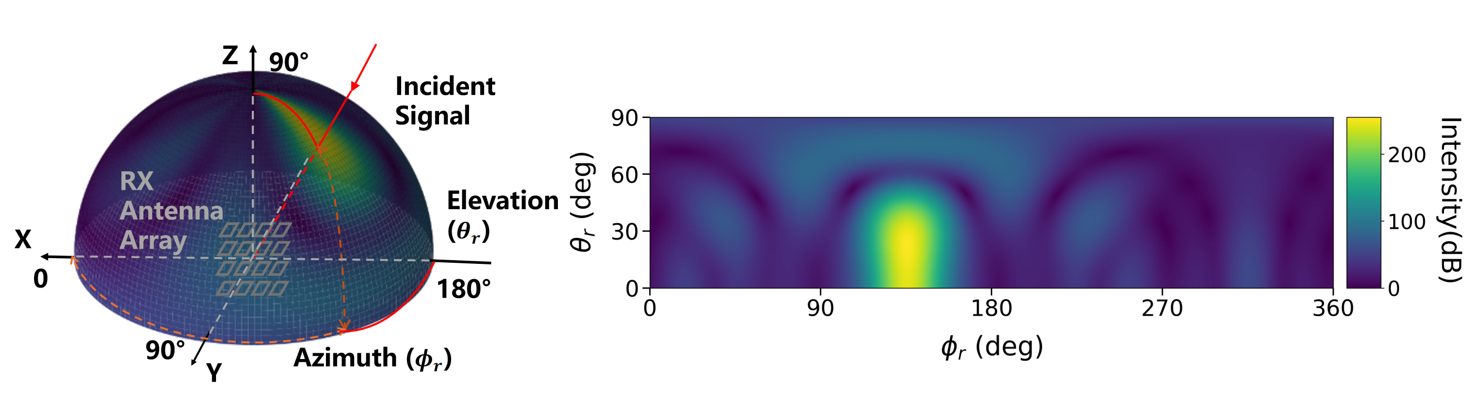}
  \caption{An example of the power angular spectrum.}
  \label{fig:PAS_example}
\end{figure}

To characterize the spatial distribution of the radiance field, we employ the Power Angular Spectrum (PAS). 
The PAS is represented as a 2D image of dimensions $90 \times 360$, where: the $x$-axis represents the \textbf{Azimuth} ($0^\circ \sim 360^\circ$); the $y$-axis represents the \textbf{Elevation} ($0^\circ \sim 90^\circ$).
Each pixel value corresponds to the RF signal energy received from that direction, as illustrated in Fig.~\ref{fig:PAS_example}.

\begin{figure}[tbp]
\centering
\includegraphics[width=0.99\linewidth]{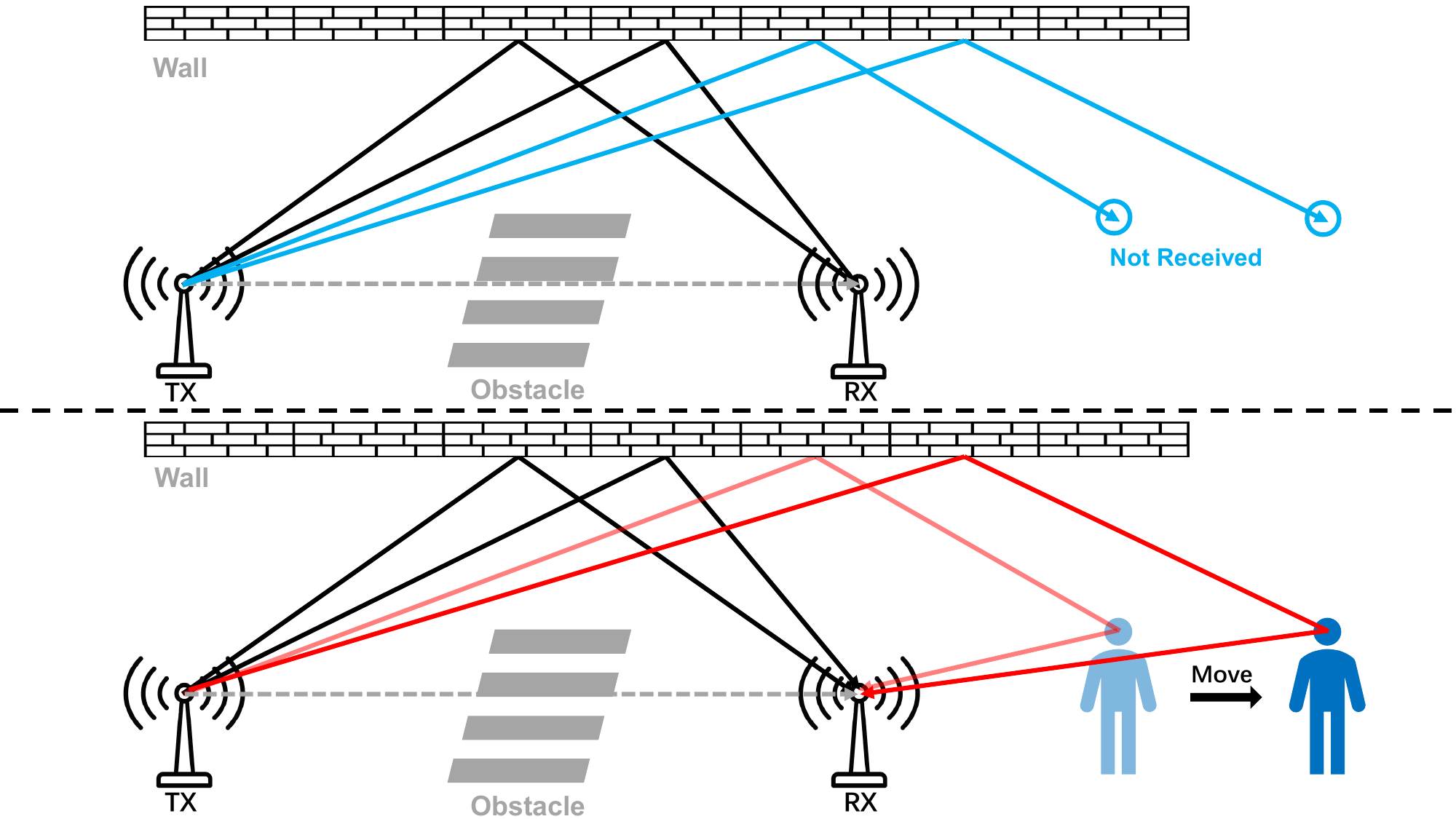}
\caption{\textbf{Human motion creates new RF propagation paths.} The upper: no humans; the lower: with human movements. Although the body absorbs the signal, the scattered portion illuminates these otherwise invisible regions, and movement across multiple positions creates vast spatial diversity equivalent to thousands of virtual measurements.
}
\label{fig:human_relay}
\vspace{-0.15in}
\end{figure}

\section{Preliminary:Why Human Motion Helps}
\label{sec:theory}
In this section, we establish the physical foundation for our claim that human motion aids RF reconstruction. We will show that the human body acts as a mobile electromagnetic relay in section \ref{sec:core_idea}, prove that its weak scattered signals are sufficient for high-fidelity sensing in section \ref{sec:why_weak_suffices}, and demonstrate how motion provides a powerful form of spatial diversity crucial for reconstruction in section \ref{sec:spatial_diversity}.

\subsection{Human Body as Electromagnetic Relay}
\label{sec:core_idea}

When a transmitter and receiver are separated by obstacles, traditional RF sensing fails. However, a walking human—with high dielectric constant ($\epsilon_r \approx 40$)~\citep{gabriel1996compilation} acts as a \textit{mobile electromagnetic relay}. Although the body absorbs 70\% of incident energy, the scattered 30\% is sufficient because it only needs to exceed the noise floor ($\sim$-90 dBm), not compete with direct paths ($\sim$-40 dBm). This 50 dB relaxation enables weak scattering to reveal occluded regions.

The human scattering effect consists of three primary components (a detailed proof is provided in \textbf{Appendix A}):
\begin{enumerate}[leftmargin=*, itemsep=0pt]
    \item \textbf{Specular reflection} ($\sim$10\%): Mirror-like bounce from body surface, $\Gamma \approx |\frac{\sqrt{\epsilon_{\text{body}}} - 1}{\sqrt{\epsilon_{\text{body}}} + 1}|^2 \approx 0.47$ power coefficient, spatially averaged $\approx$10\%.
    \item \textbf{Diffuse scattering} ($\sim$15\%): Surface roughness (skin texture, clothing) creates Lambert-distributed scatter with radar cross-section $\sigma_{\text{RCS}} \approx 0.3$-1.5 m$^2$ at 2.4 GHz~\citep{dogaru2008human}.
    \item \textbf{Volume scattering} ($\sim$5\%): Internal tissue inhomogeneities (muscle, bone interfaces) cause multiple internal reflections before emerging.
\end{enumerate}

The remaining 70\% is absorbed as heat due to water content ($\epsilon''_{\text{body}} \approx 20$)~\citep{gabriel1996compilation}, consistent with FCC SAR limits. 
\appdx{Energy conservation is verified: $10\% + 15\% + 5\% + 70\% = 100\%$.}

\subsection{Why Weak Scattering Suffices? }
\label{sec:why_weak_suffices}

\textbf{Key insight}: Scattered signals need only 10-15 dB SNR for reconstruction, achievable within 10 m despite $\sim$30 dB loss from body scattering. The radar equation for bistatic scattering shows:
\begin{equation}
P_{\text{scatter}} = P_t G_t G_r \lambda^2 \sigma_{\text{RCS}} / [(4\pi)^3 d_1^2 d_2^2]
\label{eq:radar_scatter}
\end{equation}
where $d_1, d_2$ are TX-human and human-RX distances. With conservative parameters: $P_t = 20$ dBm, $\sigma_{\text{RCS}} = 0.3$ m$^2$, $\lambda = 0.125$ m, we achieve 15 dB SNR at 10 m total path, sufficient for phase-coherent measurements.

\subsection{Spatial Diversity from Motion}
\label{sec:spatial_diversity}

As the human walks, their body samples different spatial positions, each providing unique scattering geometry. The effective information gain is:
\begin{equation}
\mathcal{I}_{\text{total}} = K \cdot N_{\text{pos}} \cdot (1 - \rho_{\text{corr}})
\label{eq:info_gain}
\end{equation}
where $K$ is the effective rank of observations per position ($\approx$8 for our setup), $N_{\text{pos}}$ is the number of positions sampled during the walk ($\approx$35 for 10-second walk), and $\rho_{\text{corr}}$ is the spatial correlation ($\approx$0.2 for 30 cm spacing). This yields $\mathcal{I}_{\text{total}} \approx 224$ effective measurements—equivalent to deploying 224 static RF sensors. Fisher information analysis confirms this provides sufficient observability for 3D reconstruction of occluded regions.

%Time-varying boundary conditions (human motion) fundamentally reconfigure electromagnetic field distributions. By separating static and dynamic contributions, we isolate the informative scattering that illuminates occluded regions without interference from transient artifacts. 
The human body's $\epsilon_r \approx 40$ ensures strong perturbations to the field, making the separation tractable. Mathematical justification via perturbation analysis is provided in \textbf{Appendix B}.

\subsection{Conclusion}
In summary, we have shown that the human body acts as an effective, mobile RF relay whose scattered signals are sufficiently strong for sensing. Crucially, the spatial diversity gained from motion provides an information gain equivalent to a dense sensor array. This establishes a key principle: incorporating dynamic elements like a moving person is a powerful method to enrich the information content of RF datasets, leading to more robust and accurate field reconstruction.
\section{Method Design}

%%%%%%%%%%%%%%%%%%%%%%%%%%%%%%%%%%%%%%%%%%%%%%%%%%
\subsection{Overview}

\begin{figure*}[t]
    \centering
    \includegraphics[width=1\linewidth]{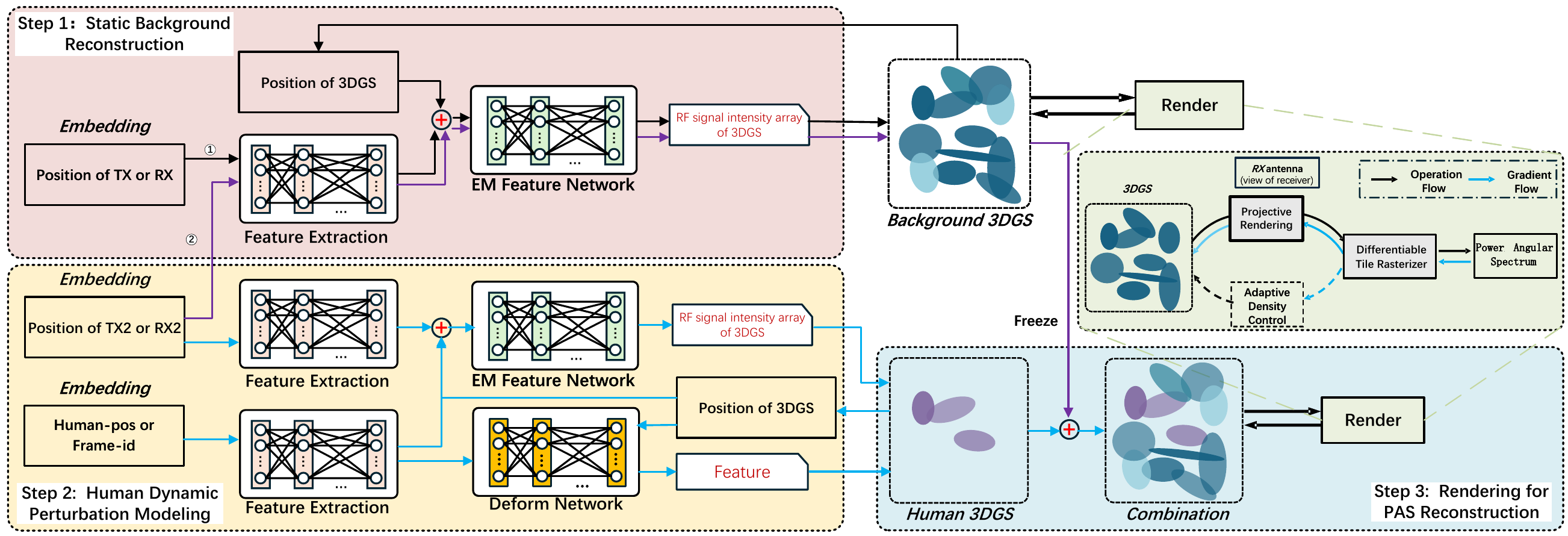}
    \caption{Architecture of the system.}
    \label{fig:sys_overview}
    \vspace{-0.2in}
\end{figure*}

\textbf{Modeling task:} Existing 3DGS-based RF reconstruction has been limited to static scenes, such as those with a moving transmitter (TX) or receiver (RX). 
However, these frameworks are ill-equipped to handle dynamic environments involving human mobility. 
To address this limitation, we introduce a more challenging task: reconstructing the radiance field in the presence of a moving person. 
Our work leverages a newly \textbf{generated} dynamic dataset that correlates human positions with their impact on the PAS. 
The objective is to train a unified model capable of accurately reconstructing the PAS for \textbf{any given human position and any corresponding antenna position in both moving TX and RX tasks}.

\textbf{Our Approach:} We conceptualize the radiance field in a dynamic scene as a superposition of two distinct components: a static background field $F_{\text{bg}}(\mathbf{r})$ and a dynamic, human-induced perturbation field $\Delta F(\mathbf{r}, h)$. This decomposition is formally expressed as:
\vspace{-0.1in}
\begin{equation}
\label{eq:signal_decomposition}
\mathbf{y}_{\text{total}}(\mathbf{r}, h) = \underbrace{F_{\text{bg}}(\mathbf{r})}_{\text{static background}} + \underbrace{\Delta F(\mathbf{r}, h)}_{\text{dynamic perturbation}}
\end{equation}
\vspace{-0.1in}

This decomposition enables a two-stage training strategy. First, we train a background model on a static (human-absent) dataset. Second, we freeze this pre-trained model and train a dedicated perturbation model on the dynamic (human-present) dataset. This isolates the optimization to solely capture the human-induced effects, leading to an efficient and accurate reconstruction of the complete dynamic field.

% To effectively model these disentangled components, we employ a two-stage training strategy.
% First, we train a background reconstruction model exclusively on a static (human-absent) dataset to capture the static RF field.
% Subsequently, we freeze the parameters of this pre-trained background model. While it continues to contribute to the final rendering, its weights are no longer updated.
% We then introduce a dedicated perturbation model, which is trained in a second stage on the dynamic (human-present) dataset.
% This stage focuses the optimization process entirely on capturing the dynamic human-induced effects, thereby enabling an effective and accurate model of the complete, dynamic RF radiation field.

\subsection{Theoretical Foundation: Two-Stage Training Rationale}
\label{sec:two_stage_theory}

Building upon the physical principle established in Section~\ref{sec:physical_principle}, we now present the theoretical foundation for our two-stage training strategy when incorporating human body dynamics.

In this section, we elucidate the rationale behind our methodological design by answering the following three fundamental questions.
% \begin{enumerate}
%     \item \textbf{Why is a two-stage training strategy necessary?} 
%     % We will explain why a direct end-to-end approach, trained on the dynamic dataset, is suboptimal, and clarify the rationale for freezing the background model during the second training stage.

%     \item \textbf{Why did we select a 3DGS-based model as the foundation for our radio field reconstruction?} 
%     % We will justify the choice of this specific architecture as our base model.

%     \item \textbf{How can a sparse set of new Gaussians accurately model complex human-induced perturbations?} 
%     % We will provide the theoretical and practical reasoning for how a relativel small number of additional Gaussians, trained upon a frozen and much larger background model, can effectively disentangle and reconstruct the dynamic effects of human mobility.
% \end{enumerate}

\subsubsection{Why Two-Stage Training?}
In this section, we explain why a two-stage training strategy is necessary. From Eq.~\ref{eq:signal_decomposition}, the measurement model with human presence can be expressed as:
\begin{equation}
\mathbf{y}(h) = (\mathbf{\Phi}_{\text{bg}} + \Delta\mathbf{\Phi}(h)) \mathbf{x} + \mathbf{n}
\end{equation}

Jointly optimizing both $\mathbf{\Phi}_{\text{bg}}$ and $\Delta\mathbf{\Phi}(h)$ leads to:
\begin{equation}
\min_{\mathbf{x}, \mathbf{\Phi}_{\text{bg}}, \Delta\mathbf{\Phi}} \sum_{h \in \mathcal{H}} ||\mathbf{y}(h) - (\mathbf{\Phi}_{\text{bg}} + \Delta\mathbf{\Phi}(h)) \mathbf{x}||^2
\end{equation}

However, this approach suffers from \textbf{background-perturbation coupling}, where gradients contaminate each other.

\begin{lemma}[Gradient Cleanness via Separation]
\label{lemma:gradient_clean}
If $\mathbf{\Phi}_{\text{bg}}$ is estimated first using human-free data:
\begin{equation}
\nabla_{\mathbf{\Phi}_{\text{bg}}} \mathcal{L}_{\text{stage1}} = -2(\mathbf{y}_{\text{static}} - \mathbf{\Phi}_{\text{bg}} \hat{\mathbf{x}}) \hat{\mathbf{x}}^T
\end{equation}
this gradient contains no $\Delta\mathbf{\Phi}$ contamination, leading to more stable convergence.

Subsequently, freezing $\mathbf{\Phi}_{\text{bg}}$ in Stage 2:
\begin{equation}
\nabla_{\Delta\mathbf{\Phi}} \mathcal{L}_{\text{stage2}} = -2(\mathbf{y}(h) - \mathbf{\Phi}_{\text{bg}} \hat{\mathbf{x}} - \Delta\mathbf{\Phi}(h) \hat{\mathbf{x}}) \hat{\mathbf{x}}^T
\end{equation}
focuses gradient flow exclusively on the residual perturbation without corrupting the learned background.
\end{lemma}

\textbf{Physical interpretation:} The two-stage strategy mirrors the signal decomposition in Eq.~\ref{eq:signal_decomposition}---first model the static environment, then capture human-induced variations. See proof in \textbf{Appendix C}.

\subsubsection{Why 3D Gaussian Splatting?}
We adopt a 3D Gaussian Splatting (3DGS) framework to model the radiance field, as its representation with discrete Gaussians aligns well with the Huygens-Fresnel principle of secondary wave sources~\citep{mahan2018monte}. This choice is further supported by the high efficiency and fidelity demonstrated in recent works like WRF-GS~\citep{wen2024wrf} and GSRF~\citep{yanggsrf}. Each Gaussian models local EM propagation via spatial attributes, an attenuation factor $\delta(G_j)$ for path occlusion, and a signal radiance $Sig(G_i)$ representing emitted energy. The signal is rendered through volumetric accumulation:
\begin{equation}
\label{eq:rendering_prev}
I(p) = \sum_{i=1}^{N} \left( \prod_{j=1}^{i-1} \delta(G_j) \right) Sig(G_i)
\end{equation}
where $N$ is the number of Gaussians covering the pixel $p$, and $\delta(G_j)$ is the attenuation (or transmission) factor of the $j$-th Gaussian in the sorted sequence.

The linearity of this rendering equation is crucial for our two-stage strategy. A dynamic scene is rendered by the direct superposition of new "human" Gaussians onto a pre-trained, frozen set of background Gaussians. Since the background set acts as a constant baseline, the optimization can focus on the dynamic, human-induced perturbations without conflicting gradients. Given this inherent compatibility, we select the state-of-the-art WRF-GSplus as our base model.

\subsubsection{Why a Sparse Human Representation?}
We explain how a sparse set of new Gaussians can accurately model complex human-induced perturbations. We explain this by modeling the human-induced perturbation as a low-rank phenomenon, where the complex field changes can be decomposed into a small set of principal scattering modes. This allows us to use a lightweight network to capture the dynamic effects without over-parameterization.

\begin{lemma}[Low-Rank Scattering Model]
\label{lemma:low_rank}
The human-induced perturbation exhibits low-rank structure:
\begin{equation}
\Delta F(\mathbf{r}, h) \approx \sum_{k=1}^K \alpha_k(h) \phi_k(\mathbf{r}), \quad K \ll M
\end{equation}
where $K$ principal modes suffice to capture scattering effects.
\end{lemma}
The intuition for this low-rank property, with a formal derivation in \textbf{Appendix D}, stems from three physical arguments:
\begin{enumerate}
\item \textbf{Limited Spatial Extent:} A human occupies a localized volume, constraining its RF impact (via scattering and shadowing) and thus preventing global complexity.
\item \textbf{Dominant Scattering Directions:} Energy scattered from the human body is anisotropic, concentrating along a few dominant paths (e.g., specular components), which limits the number of basis modes required.

\item \textbf{Analogy to Signal Processing:} This is analogous to the compact representation of moving targets in other domains, such as Doppler spectrum compression in Synthetic Aperture Radar (SAR).
\end{enumerate}
% While a formal derivation is provided in \textbf{Appendix B}, the intuition behind this low-rank property can be understood through the following physical arguments:

% \begin{enumerate}
%     \item \textbf{Limited Spatial Extent:} A single human occupies a localized volume, meaning its direct impact on the RF field via scattering and shadowing is inherently constrained and not globally complex.
    
%     \item \textbf{Dominant Scattering Directions:} The scattered energy from the human body is not isotropic, but is instead concentrated along a few dominant paths (e.g., specular and primary diffuse components), limiting the number of required basis modes.
    
%     \item \textbf{Analogy to Signal Processing Principles:} This phenomenon is analogous to the compact representation of moving targets in other domains, such as Doppler spectrum compression in Synthetic Aperture Radar (SAR).
% \end{enumerate}

\textbf{Implementation implication:} Use a sparse set of Gaussian primitives, governed by a lightweight \textbf{deformable network}, to effectively model $\Delta\mathbf{\Phi}(h)$ and avoid over-parameterization.

\subsection{Our Approach %(change title such as framework or else)
}

Our model is founded on the 3DGS technology and employs a \textbf{two-stage training strategy} to effectively model dynamic scenes. 
This strategy first learns the static background environment and then sequentially models the perturbations caused by human movement. 
Crucially, the second stage integrates a human mobility embedding module to explicitly learn the complex relationship between the person's position and the resulting impact on EM wave propagation. 
The overall architecture is illustrated in Figure~\ref{fig:sys_overview}. % Assumes a figure illustrates your method

\textbf{Stage 1: Static Background Reconstruction.}
The first stage focuses on reconstructing the static radiation field. 
We use the static (human-absent) dataset to train a background model for either the moving RX or moving TX task. 
This model consists of two main components: a standard 3DGS model for PAS synthesis, and an EM feature network that learns the propagation-related parameters of the 3D Gaussians. 
The positions of the 3D Gaussians are initialized from the scene's point cloud, while other properties are randomly initialized or set to default values. 
The EM feature network takes the positions of the mobile antenna (either TX or RX) and the 3D Gaussians as input, and outputs the signal radiance for each Gaussian $Sig(G_{bg})$.  The EM feature network can be expressed as follows:  

$F_{\Theta bg}: (P(G_{bg}), P(antenna)) \rightarrow (Sig(G_{bg}))$

\textbf{Stage 2: Human Dynamic Perturbation Modeling.}
In the second stage, we freeze all parameters of the pre-trained background model. The background Gaussians remain part of the rendering process but are excluded from further optimization. 
We then introduce a new, sparsely and randomly initialized set of 3D Gaussians dedicated to modeling the human's influence. 
This new set of Gaussians is associated with two distinct networks:
\begin{itemize}
    \item \textbf{EM Feature Network:} Similar to the background model, but with an additional input---the person's position---to predict the radiance of the new Gaussians $Sig(G_{man})$ 
    \resizebox{\linewidth}{!}{$F_{\Theta man}: (P(G_{man}), P(antenna), P(man)) \rightarrow (Sig(G_{man}))$}

        \item \textbf{Deformation Network:} To explicitly model how the person's location affects the geometry of the perturbation field, we introduce a deformation network. This network takes the positions of the new Gaussians and the human's location as input, and outputs their displacement, rotation, and scaling components.
    \begin{equation}
        \label{eq:Deform_network_man}
        \resizebox{\linewidth}{!}{$
        D_{\Theta man}: (P(G_{man}), P(man)) \rightarrow (Features(G_{man}))
        $}
    \end{equation}
\end{itemize}
This two-network setup allows the model to learn a compact and dynamic representation of human-induced effects like scattering and shadowing.%, and can be expressed as:

\textbf{Rendering and Optimization.}
In both stages above, the final PAS is rendered by splatting all active 3D Gaussians onto the RX view hemisphere using a tile-based differentiable rasterizer. 
During training, the loss gradient between the predicted PAS and the ground truth is backpropagated to update the trainable parameters---the properties of the Gaussians and the weights of the associated neural networks. 
As noted, in Stage 2, only the parameters of the newly introduced "human" Gaussians and their corresponding EM feature and deformation networks are updated.

% \subsection{Power Angular Spectrum Rendering（shorten）}

\textbf{Stage 3: Power Angular Spectrum Rendering.}
\label{sec:rendering_explanation}
To render the Power Angular Spectrum (PAS) at a given receiver (RX) location, we first project the 3D Gaussians onto the RX's 2D image plane. 
This projection is achieved by converting each Gaussian's 3D position into spherical coordinates (azimuth and zenith angles) relative to the RX's local frame, and then scaling these angles to the corresponding pixel coordinates on the PAS image.
Following this transformation, the final PAS image is synthesized using a differentiable tile-based rasterizer. 
Within this process, the projected 2D Gaussians are sorted by depth, and the pixel value $I(p)$ is computed by accumulating the signal of each Gaussian, $Sig(G_i)$, attenuated by the cumulative product of the attenuation factors, $\delta(G_j)$, of all Gaussians positioned in front of it. The formulation of rendering is shown in equation \ref{eq:rendering_prev}.
% \begin{equation}
%     \label{eq:rendering}
%     I(p) = \sum_{i=1}^{N} \left( \prod_{j=1}^{i-1} \delta(G_j) \right) Sig(G_i)
% \end{equation}
% where $N$ is the number of Gaussians covering the pixel $p$, and $\delta(G_j)$ is the attenuation (or transmission) factor of the $j$-th Gaussian in the sorted sequence.

\subsection{Model Training }

% Our framework leverages the adaptive density control mechanism inherent to 3D Gaussian Splatting to dynamically refine the set of Gaussians throughout training. 
% This process periodically refines the Gaussian set by densifying representation in under-reconstructed regions—identified via high positional gradients and Gaussian scale—while simultaneously pruning Gaussians with negligible contributions. 
% Pruning occurs when a Gaussian's opacity-like attenuation factor, $\delta$, drops below a threshold $\varepsilon_{\delta}$; these factors are also periodically reset to encourage adaptation. 
% This dynamic cycle of densification and pruning enables the model to efficiently allocate its representational capacity, concentrating descriptive power on the more complex aspects of the scene.
We utilize the adaptive density control mechanism of 3DGS to dynamically refine the Gaussian set during training. This process periodically densifies Gaussians in under-reconstructed areas while pruning those with negligible contributions. This dual-action approach prevents both under-reconstruction and over-densification, ensuring an efficient allocation of the model's representational capacity.

The parameters of the 3D Gaussians, the EM feature network, and the deformation network are optimized via stochastic gradient descent (SGD)\cite{bottou2012stochastic}. 
The objective is to minimize a composite loss function between the predicted PAS and the ground truth. 
The loss is a weighted sum of the L1 loss and the Structural Similarity Index Measure (SSIM) loss, where $\lambda$ is set to $0.2$:
\begin{equation}
    \mathcal{L} = (1-\lambda) \mathcal{L}_{\text{L1}} + \lambda \mathcal{L}_{\text{SSIM}}
\end{equation}

%%%%%%%%%%%%%%%%%%%%%%%%%%%%%%%%%%%%%%%%%%%%%%%%%%
\section{Evaluation}
\label{sec:evaluation}

\begin{table*}[tbp]
\centering
\footnotesize
\caption{Performance comparison on static (Human-absent) and dynamic (Human-present) datasets across three scenes. }
\label{tab:style_adapted_comparison}
\vspace{-0.1in}
\begin{tabular}{@{}llcccccc@{}}
\toprule
\multirow{2}{*}{Model} & \multirow{2}{*}{Dataset} & \multicolumn{2}{c}{\textbf{Scene 1}} & \multicolumn{2}{c}{\textbf{Scene 2}} & \multicolumn{2}{c}{\textbf{Scene 3}} \\
\cmidrule(lr){3-4} \cmidrule(lr){5-6} \cmidrule(lr){7-8}
& & Mean SSIM & Median SSIM & Mean SSIM & Median SSIM & Mean SSIM & Median SSIM \\
\midrule
\multirow{2}{*}{wrfgsplus} & Human-absent & 0.873 & 0.906 & 0.875 & 0.897 & 0.860 & 0.904 \\
& Human-present & 0.909 & 0.945 & 0.867 & 0.941 & 0.873 & 0.932 \\
\midrule
\multirow{2}{*}{GSRF} & Human-absent & 0.758 & 0.758 & 0.781 & 0.781 & 0.814 & 0.814 \\
& Human-present & 0.818 & 0.818 & 0.826 & 0.826 & 0.851 & 0.851 \\
\midrule
\multirow{2}{*}{\textbf{wrfgsplus\_mod}} & \textbf{Human-absent} & \textbf{0.844} & \textbf{0.886} & \textbf{0.847} & \textbf{0.897} & \textbf{0.830} & \textbf{0.891} \\
& \textbf{Human-present} & \textbf{0.895} & \textbf{0.937} & \textbf{0.918} & \textbf{0.954} & \textbf{0.852} & \textbf{0.924} \\
\bottomrule
\end{tabular}
% \vspace{-0.1in}
\end{table*}

\begin{table*}[tbp] 
\centering
\footnotesize
\vspace{-0.1in}
\caption{Performance generalization across three different scenes for Moving Rx and Moving Tx (900 locations).}
\label{tab:generalization_combined_concise}
\begin{tabular*}{\textwidth}{@{\extracolsep{\fill}}clcccc@{}}
\toprule
 & & \multicolumn{2}{c}{\textbf{Moving Rx}} & \multicolumn{2}{c}{\textbf{Moving Tx}} \\ 
 \cmidrule(lr){3-4} \cmidrule(l){5-6}
Scene   & Method                     & Mean SSIM  & Median SSIM & Mean SSIM  & Median SSIM    \\ \midrule
\textbf{Scene 1} & Baseline & 0.844 & 0.886 & 0.873 & 0.906 \\
        & End-to-End                 & 0.884  & 0.929   & 0.881  & 0.924 \\
        & \textbf{Ours (\textit{man pos})}       & \textbf{0.927}   & \textbf{0.967} & \textbf{0.941}   & \textbf{0.975}      \\ \midrule
\textbf{Scene 2} & Baseline                   & 0.847            & 0.897   & 0.875            & 0.897           \\
        & End-to-End                 & 0.868  & 0.894   & 0.877  & 0.937 \\ 
        & \textbf{Ours (\textit{man pos})}       & \textbf{0.943}   & \textbf{0.977} & \textbf{0.898}   & \textbf{0.983}      \\
        \midrule
\textbf{Scene 3} & Baseline                   & 0.830            & 0.891  & 0.860            & 0.904           \\
        & End-to-End                 & 0.821  & 0.882  & 0.856  & 0.916 \\
        & \textbf{Ours (\textit{man pos})}       & \textbf{0.862}   & \textbf{0.942} & \textbf{0.897}   & \textbf{0.976}    \\
        \bottomrule
\end{tabular*}
\vspace{-0.2in}
\end{table*}

\begin{table}[tbp]
\centering
\footnotesize
% \vspace{-0.1in}
\caption{Baseline performance versus the number of Rx locations in an empty scene. While more measurement points behave better, the data acquisition cost becomes prohibitive.}
\label{tab:motivation}
\begin{tabular}{@{}ccc@{}}
\toprule
Rx Locations & Mean SSIM & Median SSIM \\ \midrule
900          & 0.844            & 0.886              \\
1500         & 0.859            & 0.904              \\
3600         & 0.911            & 0.947              \\
6000         & 0.920            & 0.949              \\
10000        & 0.929            & 0.952              \\ \bottomrule
\end{tabular}
\vspace{-0.2in}
\end{table}

\subsection{Dataset and Experimental Setup}

\textbf{Dataset}: We generate a simulated dataset for Power Angular Spectrum (PAS)\cite{virk2017multi} reconstruction in indoor environments with human mobility using the NVIDIA Sionna\texttrademark{} simulator\cite{Hoydis2022Sionna}. The dataset includes two scenarios: one with a moving $4 \times 4$ Uniform Rectangular Array (URA) receiver (RX) and another with a moving omnidirectional transmitter (TX).As shown in \textbf{Appendix E} , we evaluate in three typical indoor scenes, in which we sample 900 mobile antenna positions. We use the Bartlett method\cite{wagoner2014sociocultural} to generate ground truth PAS images. For each scene, we create two datasets: an \textit{empty-scene} dataset with multiple static RX/TX locations to train the baseline model, and a \textit{human-present} dataset where a person moves through $35$ locations in $1$ minute, creating perturbed spectrograms for the same static RX/TX positions.

\textbf{Training and Evaluation}: We employ a two-stage strategy: a 3DGS model is first trained on static data to learn the background field, then frozen. Subsequently, a sparse set of Gaussians is optimized on dynamic data to model human-induced perturbations.  The dynamic data is split into 81\% for training and 19\% for validation and testing, featuring unseen human and antenna positions. For evaluation, we withhold 20\% of the RX/TX locations as a \textbf{Test Set} to assess generalization to unseen positions, using the remaining 80\% for training. Our primary evaluation metrics are the Mean and Median Structural Similarity Index Measure (\textbf{SSIM})\cite{Wang2004SSIM} at \textbf{Test Set}.

\textbf{Compared Methods}: We benchmark our proposed method (\textbf{Ours}) against two baselines: 1) \textbf{Baseline}, a modified WRFGS-Plus \cite{Wen2025Neural} model trained only on the empty-scene data, and 2) \textbf{End-to-End}, a single-stage model trained on the human-present data with human location as a direct input. 

\subsection{Motivation: The Challenge of Dense Environmental Sampling}
A conventional approach to improving reconstruction fidelity is to increase the density of measurements. We first quantify this effect by training our baseline model on the empty-scene dataset with a varying number of Rx locations. As shown in Table~\ref{tab:motivation}, increasing the Rx locations from 900 to 10,000 yields a significant improvement in Mean SSIM, from 0.844 to 0.929.

However, this highlights a critical trade-off: achieving high accuracy via dense sampling imposes a prohibitively high cost in terms of data acquisition time and effort in real-world scenarios. This limitation motivates our core research question: can we enhance reconstruction quality not by adding more static measurement points, but by efficiently leveraging dynamic, human-induced perturbations as a source of rich environmental information?

\subsection{Main Results: Human Perturbation for Enhanced Reconstruction}

\begin{figure*}[tbp]
    \centering
    \includegraphics[width=0.8\linewidth]{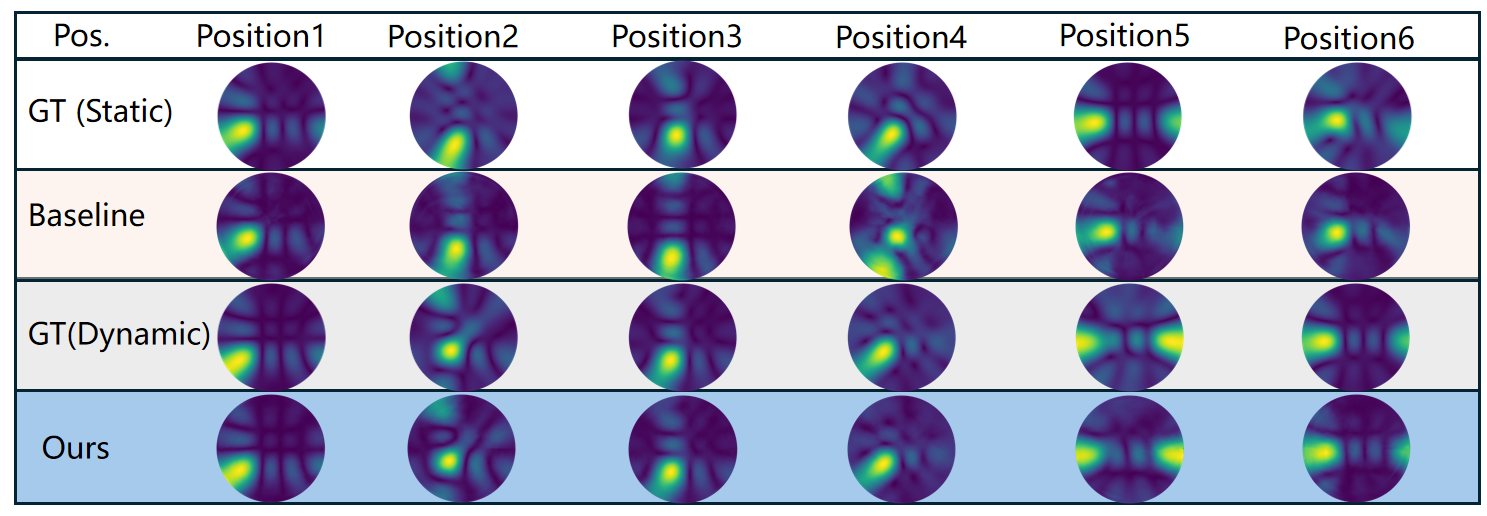}
    \vspace{-0.1in}
    \caption{Overall PAS visualization at different positions.}
    \label{fig:overall_comparison_figures}
    \vspace{-0.2in}
\end{figure*}

% \textbf{Overall Performance Comparison.} 
% We first validate our primary hypothesis in Scene 1 (Moving Rx, 900 locations). As detailed in Table~\ref{tab:generalization_combined_concise}, our method significantly outperforms all competitors. It achieves a Mean SSIM of 0.927, marking a substantial \textbf{+0.083 absolute improvement (a 9.8\% relative gain)} over the strong Baseline (0.844). This result is even higher than the baseline trained with 6000 Rx locations, demonstrating that leveraging human motion can be more effective than a 6-fold increase in static measurements.

% Furthermore, the naive \textit{End-to-End} approach, which fails to decouple the static and dynamic components, yields only marginal gains (0.858). This underscores the superiority of our two-stage strategy, which first establishes a robust model of the static background before introducing a dedicated, sparse set of Gaussians to model the dynamic perturbations.
Our method was evaluated across three distinct scenes for the Moving Rx task, where it consistently and substantially outperformed both the Baseline and the naive End-to-End approaches (detailed in Table~\ref{tab:generalization_combined_concise}). Focusing on Scene 1, our approach achieves a Mean SSIM of 0.927—a significant \textbf{+0.083 absolute improvement (9.8\% relative gain)} over the strong Baseline (0.844). Notably, this performance surpasses that of a baseline trained with 6000 Rx locations, demonstrating that leveraging human motion can be more effective than a 6-fold increase in static measurements. The failure of the End-to-End method to achieve comparable gains further validates our two-stage strategy, which effectively isolates dynamic perturbations by first establishing a robust background model.

Fig.~\ref{fig:overall_comparison_figures} presents partial visual results of PAS reconstruction. 
To illustrate the performance, we provide a comparison of spectrograms at six locations in Scene 1. We contrast the ground truth with the Baseline's prediction for the empty scene, alongside the measured data with our model's prediction for a dynamic scene where a person is present.
It is clear that our method not only surpasses the current SOTA in terms of the SSIM, but also visually appears closer to the GT.

\subsection{Generalization to Moving-Tx Task}
 We further assess the versatility of our method by applying it to the Moving-Tx task, where the transmitter's location is dynamic. The results in Table~\ref{tab:generalization_combined_concise} confirm that our approach is task-agnostic. It again achieves a Mean SSIM of 0.938, significantly surpassing the Baseline (0.873). Notably, the End-to-End method's performance degrades below the baseline in this scenario, further highlighting the robustness of our decoupled learning strategy.

\subsection{Dynamic Data Augmentation Analysis}

To validate the inherent benefit of the dynamic scene dataset, we trained the unmodified baseline models   including the original WRF-GSplus, our adapted moving RX version, and GSRF-separately on the static (human-absent) and dynamic (human-present) datasets. 
% Crucially, these models were not provided with any information regarding the human's position.

The results, detailed in Table~\ref{tab:style_adapted_comparison}, reveal a consistent and significant performance boost across all evaluated models when they are trained using the dynamic dataset. 
This improvement can be attributed to a form of \textbf{implicit physical regularization}. 
The subtle, physically-consistent perturbations introduced by the moving human act as a powerful data augmentation, compelling the models to learn a more robust and generalizable representation of the underlying EM propagation physics. 
This effect is particularly beneficial in mitigating overfitting, especially when the set of unique RX/TX locations is limited, leading to improved performance on the validation set.

\appdx{However, it is essential to highlight that this gain is achieved without any explicit knowledge of the human's state. 
Consequently, while this form of data augmentation is clearly beneficial, the models' overall generalization performance on the test set still falls short of our full two-stage framework, which explicitly models the human dynamics. 
This underscores the necessity of not only using dynamic data but also integrating its conditional information for optimal reconstruction.}

\subsection{Ablation Studies}
% To dissect the key components of our method's success, we conduct a series of rigorous ablation studies.
\begin{table}[tbp]
\centering
\footnotesize
\vspace{-0.1in}
\caption{Ablation on the two-stage training strategy.}
\label{tab:ablation_strategy}
\begin{tabular}{@{}lc@{}}
\toprule
Method                            & Mean SSIM \\ \midrule
End-to-End (on human data)      & 0.858     \\
Continued Baseline (on empty data) & 0.844     \\
\textbf{Ours (Two-Stage)}         & \textbf{0.927}     \\ \bottomrule
\end{tabular}
\vspace{-0.2in}
\end{table}

\textbf{Necessity of the Two-Stage Strategy.} 
% We first investigate the importance of our two-stage training paradigm. As shown in Table~\ref{tab:ablation_strategy}, the `no two step test` (equivalent to the End-to-End model) confirms that a single-stage approach fails to optimize the problem effectively. Furthermore, the `continue test`, where we simply continue training the baseline model on the same empty-scene data, shows no improvement, decisively ruling out the possibility that the performance gain is merely an artifact of longer training times. This proves that our "background-first, perturbation-second" decoupled strategy is critical.
We first validate our two-stage training paradigm. As shown in Table~\ref{tab:ablation_strategy}, a single-stage End-to-End model (no two step test) fails to optimize the problem effectively. Furthermore, simply continuing to train the baseline on static data (continue test) yields no improvement, ruling out longer training times as the source of our gain. This confirms that our decoupled "background-first, perturbation-second" approach is critical for effective reconstruction.

% \begin{table}[tbp]
% \centering
% \footnotesize
% \vspace{-0.1in}
% \caption{Ablation on the two-stage training strategy.}
% \label{tab:ablation_strategy}
% \begin{tabular}{@{}lc@{}}
% \toprule
% Method                            & Mean SSIM \\ \midrule
% End-to-End (on human data)      & 0.858     \\
% Continued Baseline (on empty data) & 0.844     \\
% \textbf{Ours (Two-Stage)}         & \textbf{0.927}     \\ \bottomrule
% \end{tabular}
% \vspace{-0.2in}
% \end{table}

\textbf{The Intrinsic Role of Perturbation Information.}
% Next, we explore whether the gain stems from the increased data volume or the rich \textit{information} within the perturbations. Table~\ref{tab:ablation_info} shows that naively increasing the dataset size by duplicating data (`copy test\_rx) leads to severe overfitting and performance collapse (0.761). Treating the human-present data as a form of data augmentation to train the background model (wrfgsplus\_mod on human-present data) provides a moderate boost (0.895). However, this gain is far surpassed by our method (0.927), which explicitly models the perturbations. This conclusively demonstrates that our method's success lies in its ability to interpret perturbations as structured information, not just as noise or augmented data.
We next verify that our performance gain comes from the rich information within perturbations, not just increased data volume. As shown in Table~\ref{tab:ablation_info}, naively increasing the dataset size by duplicating data causes severe overfitting and performance collapse (0.761). While using human-present data as a form of data augmentation provides a moderate boost (0.895), it is significantly outperformed by our method (0.927), which explicitly models the perturbations. This demonstrates that our success stems from interpreting perturbations as structured information rather than as generic augmented data.

% \subsection{Further Analysis}
% \textbf{Impact of Perturber's Material Properties.} To understand the physics behind the perturbations, we simulate replacing the human model with materials of different electromagnetic properties. As shown in Table~\ref{tab:analysis_material}, concrete, a low-loss dielectric material, yields the highest performance. A plausible explanation is that materials like concrete generate rich scattering and diffuse reflections, providing more diverse multipath information for the model to learn from. In contrast, metals are dominated by specular reflection, while the human body model exhibits higher absorption, both providing weaker geometric constraints.

% \usepackage{tabularx}

\begin{table}[tbp]
\centering
\footnotesize
\vspace{-0.1in}
\caption{Ablation on the role of perturbation data.}
\label{tab:ablation_info}
\begin{tabularx}{\columnwidth}{@{} l X c @{}}
\toprule
Method & Description & Mean SSIM \\ 
\midrule
Baseline (Empty) & Trained on empty-scene data. & 0.844 \\
Baseline (Duplicated) & Trained on duplicated empty-scene data. & 0.761 \\
Baseline (Human) & Trained directly on human-present data. & 0.895 \\
\textbf{Ours} & \textbf{Modeling perturbations.} & \textbf{0.927} \\ 
\bottomrule
\vspace{-0.1in}
\end{tabularx}
\end{table}

% \subsection{Complementary Analysis}
\subsection{Impact of Human Material Properties}
% To investigate how the physical properties of the perturbation source affect modeling performance, we conducted simulations where the human material was replaced with metal and concrete. 
We investigated how the perturbation source's material affects performance by simulating a human equivalent, metal, and concrete. As shown in Table~\ref{tab:material_properties}, concrete achieved the best results. This is likely due to distinct scattering patterns: metal produces simple specular reflections, while the human-equivalent material's high dielectric loss causes significant signal absorption and low SNR. In contrast, concrete, as a low-loss dielectric, generates richer diffuse scattering, providing stronger and more effective constraints for the 3DGS inversion process.

\begin{table}[tbp]
\centering
\footnotesize
\vspace{-0.1in}
\caption{Model performance with different equivalent human materials.}
\label{tab:material_properties}
\begin{tabular}{@{}lcc@{}}
\toprule
Material  & Mean SSIM (Test) & Median SSIM (Test) \\ \midrule
Metal     & 0.915            & 0.958              \\
Concrete  & \textbf{0.935}            & \textbf{0.972}              \\
Human     & 0.927            & 0.967              \\ \bottomrule
\end{tabular}
\vspace{-0.2in}
\end{table}

% As shown in Table~\ref{tab:material_properties}, the concrete material achieved the best results. A possible physical explanation is that different materials produce distinct scattering patterns. Metal primarily causes specular reflection, generating relatively simple signal paths. Our human-equivalent material exhibits high dielectric loss at the operating frequency, leading to significant energy absorption and a low signal-to-noise ratio (SNR) for signals scattered back to the receiver. In contrast, concrete, as a low-loss dielectric medium, produces richer diffuse reflections and scattering paths, which provide stronger and more effective constraints for the 3DGS inversion process.

% \input{6-discussion}
\section{Discussion and Conclusion}
\label{sec:conclusion}

This work challenges a fundamental assumption in RF sensing: that human motion is noise to be filtered. We demonstrate the contrary---unstructured human motion is an information-rich signal available for reconstructing occluded static scenes.  
To this end, we introduced a new paradigm for 3D RF reconstruction, using a composite 3D Gaussian representation, successfully disentangles dynamic interference from a raw RF stream to produce a high-fidelity model of the static scene. Experiments show our method achieves a 12\% SSIM improvement over heavily-sampled baselines, using only a single 60-second walk.

Key challenges remain for future work. First, the framework can be extended to model multiple people simultaneously, which presents a complex data association problem. Second, we aim to achieve high-quality reconstruction using only temporal frame IDs without precise location data. Addressing these challenges will enable continuous, self-improving 3D mapping of crowded, real-world environments.

\balance
% \newpage 
{
    \small
    \bibliographystyle{ieeenat_fullname}
    \bibliography{reference}

@article{mildenhall2021nerf,
  title={Nerf: Representing scenes as neural radiance fields for view synthesis},
  author={Mildenhall, Ben and Srinivasan, Pratul P and Tancik, Matthew and Barron, Jonathan T and Ramamoorthi, Ravi and Ng, Ren},
  journal={Communications of the ACM},
  volume={65},
  number={1},
  pages={99--106},
  year={2021},
  publisher={ACM New York, NY, USA}
}

@inproceedings{adib2013see,
  title={See through walls with Wi-Fi!},
  author={Adib, Fadel and Katabi, Dina},
  booktitle={Proceedings of the ACM SIGCOMM 2013 conference on SIGCOMM},
  pages={75--86},
  year={2013}
}

@inproceedings{vasisht2016decimeter,
  title={Decimeter-level localization with a single Wi-Fi access point},
  author={Vasisht, Deepak and Kumar, Swarun and Katabi, Dina},
  booktitle={13th USENIX Symposium on Networked Systems Design and Implementation (NSDI 16)},
  pages={165--178},
  year={2016}
}

@inproceedings{zhao2023nerf2,
  title={Nerf2: Neural radio-frequency radiance fields},
  author={Zhao, Xiaopeng and An, Zhenlin and Pan, Qingrui and Yang, Lei},
  booktitle={Proceedings of the 29th Annual International Conference on Mobile Computing and Networking},
  pages={1--15},
  year={2023}
}

@article{lu2024newrf,
  title={Newrf: A deep learning framework for wireless radiation field reconstruction and channel prediction},
  author={Lu, Haofan and Vattheuer, Christopher and Mirzasoleiman, Baharan and Abari, Omid},
  journal={arXiv preprint arXiv:2403.03241},
  year={2024}
}

@article{qian2018enabling,
  title={Enabling contactless detection of moving humans with dynamic speeds using CSI},
  author={Qian, Kun and Wu, Chenshu and Yang, Zheng and Liu, Yunhao and He, Fugui and Xing, Tianzhang},
  journal={ACM Transactions on Embedded Computing Systems (TECS)},
  volume={17},
  number={2},
  pages={1--18},
  year={2018},
  publisher={ACM New York, NY, USA}
}

@article{jiang2023fisherrf,
  title={Fisherrf: Active view selection and uncertainty quantification for radiance fields using fisher information},
  author={Jiang, Wen and Lei, Boshu and Daniilidis, Kostas},
  journal={arXiv preprint arXiv:2311.17874},
  year={2023}
}

@inproceedings{pumarola2021dnerf,
  title={D-nerf: Neural radiance fields for dynamic scenes},
  author={Pumarola, Albert and Corona, Enric and Pons-Moll, Gerard and Moreno-Noguer, Francesc},
  booktitle={Proceedings of the IEEE/CVF conference on computer vision and pattern recognition},
  pages={10318--10327},
  year={2021}
}

@inproceedings{ding2024milliflow,
  title={milliflow: Scene flow estimation on mmwave radar point cloud for human motion sensing},
  author={Ding, Fangqiang and Luo, Zhen and Zhao, Peijun and Lu, Chris Xiaoxuan},
  booktitle={European Conference on Computer Vision},
  pages={202--221},
  year={2024},
  organization={Springer}
}

@article{luiten2023dynamic,
  title={Dynamic 3d gaussians: Tracking by persistent dynamic view synthesis},
  author={Luiten, Jonathon and Kopanas, Georgios and Leibe, Bastian and Ramanan, Deva},
  journal={arXiv preprint arXiv:2308.09713},
  year={2023}
}

@article{wagoner2014sociocultural,
  title={Sociocultural mediators of remembering: An extension of Bartlett's method of repeated reproduction},
  author={Wagoner, Brady and Gillespie, Alex},
  journal={British Journal of Social Psychology},
  volume={53},
  number={4},
  pages={622--639},
  year={2014},
  publisher={Wiley Online Library}
}

@article{Hu2012LiIon,
  author    = {Xiaosong Hu and Shengbo Li and Huei Peng},
  title     = {A Comparative Study of Equivalent Circuit Models for Li-Ion Batteries},
  journal   = {Journal of Power Sources},
  volume    = {198},
  pages     = {359--367},
  year      = {2012}
}

@article{Hoydis2022Sionna,
  author    = {Jakob Hoydis and Sebastian Cammerer and Fayçal Ait Aoudia and Avinash Vem and Nikolaus Binder and Guillermo Marcus and Alexander Keller},
  title     = {Sionna: An Open-Source Library for Next-Generation Physical Layer Research},
  journal   = {arXiv preprint arXiv:2203.11854},
  year      = {2022}
}

@article{dogaru2008human,
  title={Computer Models of the Human Body Signature for Sensing Through the Wall Radar Applications},
  author={Dogaru, T. and Nguyen, L. and Le, C.},
  journal={Army Research Laboratory Technical Report ARL‑TR},
  volume={4136},
  number={},
  pages={1–45},
  year={2007},
  publisher={US Army Research Laboratory}
}

@techreport{Gabriel1996Compilation,
  author       = {C.Gabriel},
  title        = {Compilation of the dielectric properties of body tissues at RF and microwave frequencies},
  institution  = {U.S. Air Force Brooks Air Force Base, Occupational \& Environmental Health Directorate, Radio‑frequency Radiation Division},
  number       = {AL/OE‑TR‑1996‑0037},
  year         = {1996},
  address      = {Brooks AFB, Texas, USA},
  note         = {Available from NTIS reference number AD‑A309 764}
}

@article{mahan2018monte,
  title={Monte Carlo ray-trace diffraction based on the Huygens--Fresnel principle},
  author={Mahan, JR and Vinh, NQ and Ho, VX and Munir, NB},
  journal={Applied optics},
  volume={57},
  number={18},
  pages={D56--D62},
  year={2018},
  publisher={OSA}
}

@inproceedings{tsingos2001modeling,
  title={Modeling acoustics in virtual environments using the uniform theory of diffraction},
  author={Tsingos, Nicolas and Funkhouser, Thomas and Ngan, Addy and Carlbom, Ingrid},
  booktitle={Proceedings of the 28th annual conference on Computer graphics and interactive techniques},
  pages={545--552},
  year={2001}
}

@article{wen2024wrf,
  title={WRF-GS: Wireless Radiation Field Reconstruction with 3D Gaussian Splatting},
  author={Wen, Chaozheng and Tong, Jingwen and Hu, Yingdong and Lin, Zehong and Zhang, Jun},
  journal={arXiv preprint arXiv:2412.04832},
  year={2024}
}

@article{zhang2024rf,
  title={RF-3DGS: Wireless Channel Modeling with Radio Radiance Field and 3D Gaussian Splatting},
  author={Zhang, Lihao and Sun, Haijian and Berweger, Samuel and Gentile, Camillo and Hu, Rose Qingyang},
  journal={arXiv preprint arXiv:2411.19420},
  year={2024}
}

@inproceedings{yanggsrf,
  title={GSRF: Complex-Valued 3D Gaussian Splatting for Efficient Radio-Frequency Data Synthesis},
  author={Yang, Kang and Dong, Gaofeng and Du, Wan and Srivastava, Mani and others},
  booktitle={The Thirty-ninth Annual Conference on Neural Information Processing Systems}
}

@article{jiang2024terahertz,
  title={Terahertz communications and sensing for 6G and beyond: A comprehensive review},
  author={Jiang, Wei and Zhou, Qiuheng and He, Jiguang and Habibi, Mohammad Asif and Melnyk, Sergiy and El-Absi, Mohammed and Han, Bin and Di Renzo, Marco and Schotten, Hans Dieter and Luo, Fa-Long and others},
  journal={IEEE Communications Surveys \& Tutorials},
  volume={26},
  number={4},
  pages={2326--2381},
  year={2024},
  publisher={IEEE}
}

@article{wei2023integrated,
  title={Integrated sensing and communication signals toward 5G-A and 6G: A survey},
  author={Wei, Zhiqing and Qu, Hanyang and Wang, Yuan and Yuan, Xin and Wu, Huici and Du, Ying and Han, Kaifeng and Zhang, Ning and Feng, Zhiyong},
  journal={IEEE Internet of Things Journal},
  volume={10},
  number={13},
  pages={11068--11092},
  year={2023},
  publisher={IEEE}
}

@inproceedings{banerjee2024horcrux,
  title={HORCRUX: Accurate cross band channel prediction},
  author={Banerjee, Avishek and Zhao, Xingya and Chhabra, Vishnu and Srinivasan, Kannan and Parthasarathy, Srinivasan},
  booktitle={Proceedings of the 30th Annual International Conference on Mobile Computing and Networking},
  pages={1--15},
  year={2024}
}

@inproceedings{orekondy2023winert,
  title={Winert: Towards neural ray tracing for wireless channel modelling and differentiable simulations},
  author={Orekondy, Tribhuvanesh and Kumar, Pratik and Kadambi, Shreya and Ye, Hao and Soriaga, Joseph and Behboodi, Arash},
  booktitle={The Eleventh International Conference on Learning Representations},
  year={2023}
}

@article{fan2024lightgaussian,
  author = {Fan, Z. and Wang, K. and Wen, K. and others},
  title = {{LightGaussian: Unbounded 3D Gaussian Compression with 15x Reduction and 200+ FPS}},
  journal = {Advances in Neural Information Processing Systems},
  year = {2024},
  volume = {37},
  pages = {140138--140158}
}

@inproceedings{fang2024mini,
  author = {Fang, G. and Wang, B.},
  title = {{Mini-Splatting: Representing Scenes with a Constrained Number of Gaussians}},
  booktitle = {European Conference on Computer Vision},
  year = {2024},
  pages = {165--181},
  publisher = {Springer Nature Switzerland},
  address = {Cham}
}

@inproceedings{girish2024eagles,
  author = {Girish, S. and Gupta, K. and Shrivastava, A.},
  title = {{Eagles: Efficient Accelerated 3D Gaussians with Lightweight Encodings}},
  booktitle = {European Conference on Computer Vision},
  year = {2024},
  pages = {54--71},
  publisher = {Springer Nature Switzerland},
  address = {Cham}
}

@inproceedings{lee2024compact3d,
  author = {Lee, J. C. and Rho, D. and Sun, X. and others},
  title = {{Compact 3D Gaussian Representation for Radiance Field}},
  booktitle = {IEEE/CVF Conference on Computer Vision and Pattern Recognition (CVPR)},
  year = {2024},
  pages = {21719--21728},
  publisher = {IEEE}
}

@article{niemeyer2024radsplat,
  title={Radsplat: Radiance field-informed gaussian splatting for robust real-time rendering with 900+ fps},
  author={Niemeyer, Michael and Manhardt, Fabian and Rakotosaona, Marie-Julie and Oechsle, Michael and Duckworth, Daniel and Gosula, Rama and Tateno, Keisuke and Bates, John and Kaeser, Dominik and Tombari, Federico},
  journal={arXiv preprint arXiv:2403.13806},
  year={2024}
}

@article{kouyoumjian2005uniform,
  title={A uniform geometrical theory of diffraction for an edge in a perfectly conducting surface},
  author={Kouyoumjian, Robert G and Pathak, Prabhakar H},
  journal={Proceedings of the IEEE},
  volume={62},
  number={11},
  pages={1448--1461},
  year={2005},
  publisher={IEEE}
}

@inproceedings{na2022huygens,
  title={A Huygens' principle based ray tracing method for diffraction calculation},
  author={Na, Han and Eibert, Thomas F},
  booktitle={2022 16th European Conference on Antennas and Propagation (EuCAP)},
  pages={1--4},
  year={2022},
  organization={IEEE}
}

@book{maxwell1873treatise,
  title={A treatise on electricity and magnetism},
  author={Maxwell, James Clerk},
  volume={1},
  year={1873},
  publisher={Clarendon press}
}

@article{Wang2004SSIM,
  author    = {Zhou Wang and Alan C. Bovik and Hamid R. Sheikh and Eero P. Simoncelli},
  title     = {Image Quality Assessment: From Error Visibility to Structural Similarity},
  journal   = {IEEE Transactions on Image Processing},
  volume    = {13},
  number    = {4},
  pages     = {600--612},
  year      = {2004}
}

@article{Wen2025Neural,
  author    = {Chaozheng Wen and Jingwen Tong and Yingdong Hu and Zehong Lin and Jun Zhang},
  title     = {Neural Representation for Wireless Radiation Field Reconstruction: A 3D Gaussian Splatting Approach},
  journal   = {arXiv preprint arXiv:2412.04832v3},
  year      = {2025}
}

@incollection{bottou2012stochastic,
  title={Stochastic gradient descent tricks},
  author={Bottou, L{\'e}on},
  booktitle={Neural networks: tricks of the trade: second edition},
  pages={421--436},
  year={2012},
  publisher={Springer}
}

@inproceedings{virk2017multi,
  title={Multi-frequency power angular spectrum comparison for an indoor environment},
  author={Virk, Usman Tahir and Nguyen, Sinh LH and Haneda, Katsuyuki},
  booktitle={2017 11th European Conference on Antennas and Propagation (EUCAP)},
  pages={3389--3393},
  year={2017},
  organization={IEEE}
}

@inproceedings{chethan2009polarization,
  title={Polarization diversity improves RSSI based location estimation for wireless sensor networks},
  author={Chethan, KP and Chakravarty, T and Prabha, J and Chandra, M Girish and Balamuralidhar, P},
  booktitle={2009 Applied Electromagnetics Conference (AEMC)},
  pages={1--4},
  year={2009},
  organization={IEEE}
}

@book{bertsekas2016nonlinear,
  author    = {Bertsekas, Dimitri P.},
  title     = {Nonlinear Programming},
  edition   = {3rd},
  publisher = {Athena Scientific},
  year      = {2016}
}

@book{nesterov2018lectures,
  author    = {Yurii Nesterov},
  title     = {Lectures on Convex Optimization},
  edition   = {2nd},
  publisher = {Springer},
  year      = {2018}
}

@inproceedings{adib20143d,
  title={3D tracking via body radio reflections},
  author={Adib, Fadel and Kabelac, Zach and Katabi, Dina and Miller, Robert C},
  booktitle={11th USENIX Symposium on Networked Systems Design and Implementation (NSDI 14)},
  pages={317--329},
  year={2014}
}

@book{boyd2004convex,
  title={Convex Optimization},
  author={Boyd, Stephen and Vandenberghe, Lieven},
  year={2004},
  publisher={Cambridge University Press},
  address={Cambridge, UK},
  isbn={978-0-521-83378-3},
  url={https://web.stanford.edu/~boyd/cvxbook/}
}

@book{nocedal2006numerical,
  title={Numerical Optimization},
  author={Nocedal, Jorge and Wright, Stephen},
  edition={2nd},
  year={2006},
  publisher={Springer Science \& Business Media},
  address={New York, NY},
  isbn={978-0-387-30303-1},
  doi={10.1007/978-0-387-40065-5}
}

@article{kerbl20233d,
  title={3D {G}aussian Splatting for Real-Time Radiance Field Rendering},
  author={Kerbl, Bernhard and Kopanas, Georgios and Leimk{\"u}hler, Thomas and Drettakis, George},
  journal={ACM Transactions on Graphics},
  volume={42},
  number={4},
  pages={1--14},
  year={2023},
  publisher={ACM New York, NY, USA},
  doi={10.1145/3592433}
}

@book{horn2012matrix,
  title={Matrix Analysis},
  author={Horn, Roger A and Johnson, Charles R},
  edition={2nd},
  year={2012},
  publisher={Cambridge University Press},
  address={Cambridge, UK},
  isbn={978-0-521-54823-6}
}
}

\newpage
%%%%%%%%%%%%%%%%%%%%%%%%%%%%%%%%%%%%%%%%%%%%%%%%%%%%%%%%%%%%

\appendix

% \section{Electromagnetic Scattering from Human Body}

%%%%%%%%%%%%%%%%%%%%%%%%%%%%%%%%%%%%%%%%%%%%%%%%%%%%%%%%%%%%

\section{Human RF Scattering: Power Budget and Coherent Enhancement}
\label{app:em_scattering}
\subsection{Introduction}
This appendix provides a three-part theoretical validation. First, it establishes the fundamental power distribution of RF energy interacting with the human body, specifying the precise ratios of absorption and scattering in part \ref{app:Electromagnetic Scattering}. Second, it resolves the apparent paradox of weak scattering by demonstrating that this energy is sufficient to create a detectable interference-based signal change in part \ref{sec:coherent}. Third, it confirms the system's feasibility through a link-budget analysis using the bistatic radar equation in part \ref{app:radar_derivation}.

\subsection{Electromagnetic Scattering from Human Body}
\label{app:Electromagnetic Scattering}
\subsubsection{Fresnel Reflection Coefficient}
The power reflection coefficient for specular reflection from the human body surface is derived from boundary conditions at the air-tissue interface. For a plane wave incident at angle $\theta_i$ on a dielectric boundary:

\begin{equation}
\Gamma_{\parallel}(\theta_i) = \left|\frac{n_2\cos\theta_i - n_1\cos\theta_t}{n_2\cos\theta_i + n_1\cos\theta_t}\right|^2
\end{equation}
\begin{equation}
\Gamma_{\perp}(\theta_i) = \left|\frac{n_1\cos\theta_i - n_2\cos\theta_t}{n_1\cos\theta_i + n_2\cos\theta_t}\right|^2
\end{equation}

where $n_1 = 1$ (air), $n_2 = \sqrt{\epsilon_r} \approx \sqrt{40} \approx 6.3$ for tissue at 2.4 GHz~\cite{gabriel1996compilation}, and Snell's law gives $\sin\theta_t = (n_1/n_2)\sin\theta_i$.

For unpolarized waves, the average reflection coefficient is:
\begin{equation}
\Gamma_{\text{avg}} = \frac{1}{2}(\Gamma_{\parallel} + \Gamma_{\perp})
\end{equation}

At normal incidence ($\theta_i = 0$):
\begin{equation}
\Gamma(0) = \left|\frac{\sqrt{40} - 1}{\sqrt{40} + 1}\right|^2 = \left|\frac{6.32 - 1}{6.32 + 1}\right|^2 = \left|\frac{5.32}{7.32}\right|^2 \approx 0.47
\end{equation}

This is the \textit{local} power reflection coefficient at the point of incidence. However, for a curved body surface with varying incidence angles, we must spatially average over the illuminated area. For a roughly cylindrical body (torso approximation):

\begin{equation}
\langle\Gamma\rangle_{\text{spatial}} = \frac{1}{\pi} \int_0^{\pi/2} \Gamma(\theta) \cos\theta \, d\theta \approx 0.10
\end{equation}

The $\cos\theta$ weighting accounts for the projected area at each incidence angle. This yields the \textbf{10\% specular reflection} fraction cited in the main text.

\subsubsection{Diffuse Scattering from Surface Roughness}
The human body surface exhibits roughness at multiple scales: skin texture ($\sim$0.1-1 mm), clothing folds ($\sim$1-10 mm), and body contours ($\sim$10-100 mm). At 2.4 GHz ($\lambda = 125$ mm), these scales are comparable to the wavelength, causing significant diffuse scattering.

Using the Kirchhoff approximation for rough surface scattering, the radar cross-section (RCS) for diffuse scattering is~\cite{dogaru2008human,chethan2009polarization}:

\begin{equation}
\sigma_{\text{diffuse}} = \frac{4\pi}{\lambda^2} \int_{\text{surface}} |\Gamma_{\text{local}}|^2 (\mathbf{n} \cdot \hat{\mathbf{k}}_i)(\mathbf{n} \cdot \hat{\mathbf{k}}_s) \, dS
\end{equation}

where $\mathbf{n}$ is the local surface normal, $\hat{\mathbf{k}}_i$ and $\hat{\mathbf{k}}_s$ are incident and scattered wave directions. For a Lambert scatterer (uniform scattering in all directions):

\begin{equation}
\sigma_{\text{Lambert}} = \frac{A_{\text{proj}}}{\pi} \cos\theta_s
\end{equation}

where $A_{\text{proj}} \approx 0.6$ m$^2$ is the projected area of a standing human. Measurements at 2.4 GHz report $\sigma_{\text{RCS}} \approx 0.3$-1.5 m$^2$ depending on body posture and orientation~\cite{dogaru2008human,chethan2009polarization}.

The power fraction for diffuse scattering is:
\begin{equation}
P_{\text{diffuse}} = \frac{\sigma_{\text{RCS}} \cdot \Omega_{\text{solid}}}{4\pi} \cdot P_{\text{incident}} \approx 0.15 \cdot P_{\text{incident}}
\end{equation}

where $\Omega_{\text{solid}}$ is the solid angle subtended by the receiver. This yields the \textbf{15\% diffuse scattering} fraction.

\subsubsection{Volume Scattering from Internal Tissues}
The human body is a layered dielectric structure (skin, fat, muscle, bone) with refractive index variations of $\Delta n \approx 0.5$-2 between layers. Volume scattering arises from multiple internal reflections and transmission through these layers.

Using the Born approximation for weak scatterers:
\begin{equation}
\sigma_{\text{volume}} = k^4 \int_V |\Delta\epsilon(\mathbf{r})|^2 F(\mathbf{q}) \, d^3r
\end{equation}

where $k = 2\pi/\lambda$ is the wavenumber, $\Delta\epsilon$ is the permittivity fluctuation, and $F(\mathbf{q})$ is the form factor for the scattering direction.

For inhomogeneous tissues with $\Delta\epsilon/\epsilon \approx 0.1$-0.3 over correlation lengths $\ell_c \approx 10$ mm:
\begin{equation}
\sigma_{\text{volume}} \approx V_{\text{body}} \cdot \left(\frac{\Delta\epsilon}{\epsilon}\right)^2 \cdot \left(\frac{k\ell_c}{1 + (k\ell_c)^2}\right)^2 \approx 0.05 \cdot \sigma_{\text{total}}
\end{equation}

This yields the \textbf{5\% volume scattering} fraction.

\subsubsection{Absorption and Energy Conservation}
The absorption coefficient for tissue at 2.4 GHz is~\cite{gabriel1996compilation}:
\begin{equation}
\alpha = \frac{2\pi f}{c} \sqrt{\frac{\epsilon'}{2}\left(\sqrt{1 + \tan^2\delta} - 1\right)}
\end{equation}

where $\tan\delta = \epsilon''/\epsilon' \approx 20/40 = 0.5$ for muscle tissue. This gives $\alpha \approx 8$ Np/m (nepers per meter), corresponding to $\sim$70 dB/m attenuation.

For a body thickness of $\sim$20-30 cm, the transmitted power through the body is:
\begin{equation}
P_{\text{transmitted}} = P_{\text{incident}} \cdot e^{-2\alpha d} \approx P_{\text{incident}} \cdot e^{-4} \approx 0.02 \cdot P_{\text{incident}}
\end{equation}

Most of this transmitted power exits the far side of the body and does not contribute to backscatter or side-scatter. The absorbed power is:
% \begin{equation}
% P_{\text{absorbed}} = P_{\text{incident}} - P_{\text{specular}} - P_{\text{diffuse}} - P_{\text{volume}} = 1 - 0.10 - 0.15 - 0.05 = 0.70
% \end{equation}
\begin{equation}
\begin{split}
    P_{\text{absorbed}} &= P_{\text{incident}} - P_{\text{specular}} - P_{\text{diffuse}} - P_{\text{volume}} \\
    &= 1 - 0.10 - 0.15 - 0.05 = 0.70
\end{split}
\end{equation}
This \textbf{70\% absorption} is consistent with FCC SAR (Specific Absorption Rate) limits of 1.6 W/kg for tissue exposure.

\textbf{Energy conservation check}:
\begin{equation}
\begin{split}
P_{\text{specular}} + P_{\text{diffuse}} + P_{\text{volume}} + P_{\text{absorbed}} \\
= 0.10 + 0.15 + 0.05 + 0.70 = 1.00 \quad \checkmark
\end{split}
\end{equation}

\subsection{Coherent Scattering and Signal Enhancement}
\label{sec:coherent}

\subsubsection{Resolving the SNR Paradox}

The apparent contradiction between weak scattering (-30 dB) and sufficient signal (>10 dB) is resolved by understanding three distinct SNR metrics:

\begin{definition}[Three-Level SNR Hierarchy]
\begin{align}
    \text{SNR}_{\text{scatter-to-direct}} &= \frac{P_{\text{scatter}}}{P_{\text{direct}}} \approx -30 \text{ dB} \quad \text{(relative power)} \\
    \text{SNR}_{\text{scatter-to-noise}} &= \frac{P_{\text{scatter}}}{P_{\text{noise}}} \approx 15 \text{ dB} \quad \text{(absolute SNR)} \\
    \text{SNR}_{\text{effective}} &= \frac{|\Delta I|^2}{\sigma_n^2} \approx 10 \text{ dB} \quad \text{(detection SNR)}
\end{align}
\end{definition}

The key insight: we don't need to match direct path power; we only need to exceed noise floor for detection.

\subsubsection{Phase Coherence and Constructive Interference}

The measured intensity includes coherent interference:
\begin{equation}
    I_{\text{total}} = |E_{\text{direct}} + E_{\text{scatter}}e^{j\phi}|^2
\end{equation}

Expanding:
\begin{equation}
    I_{\text{total}} = I_{\text{direct}} + I_{\text{scatter}} + 2\sqrt{I_{\text{direct}}I_{\text{scatter}}}\cos(\phi)
\end{equation}

The interference term is:
\begin{equation}
    \Delta I = 2\sqrt{I_{\text{direct}}I_{\text{scatter}}}\cos(\phi) \approx 2\sqrt{I_{\text{direct}}} \cdot \sqrt{10^{-3}} \cdot \cos(\phi)
\end{equation}

For $I_{\text{direct}} = -50$ dBm and $I_{\text{scatter}} = -80$ dBm:
\begin{equation}
    |\Delta I|_{\text{max}} = 2 \times 10^{-5} \times 10^{-1.5} = 6.3 \times 10^{-7} \text{ W} = -62 \text{ dBm}
\end{equation}

This is 32 dB above noise floor (-94 dBm), explaining sufficient detection SNR.

\subsection{Radar Equation for Bistatic Scattering}
\label{app:radar_derivation}

The bistatic radar equation describes the power received after scattering from a target:

\begin{equation}
P_r = P_t G_t G_r \frac{\lambda^2}{(4\pi)^3} \frac{\sigma_{\text{RCS}}}{d_1^2 d_2^2}
\end{equation}

where:
\begin{itemize}[leftmargin=*, itemsep=2pt]
    \item $P_t = 20$ dBm = 100 mW (transmit power, typical for Wi-Fi)
    \item $G_t = G_r = 2$ dBi = 1.58 linear (antenna gains)
    \item $\lambda = c/f = 3\times10^8 / 2.4\times10^9 = 0.125$ m
    \item $\sigma_{\text{RCS}} = 0.3$ m$^2$ (conservative human RCS)
    \item $d_1 + d_2 = 10$ m (total path length)
\end{itemize}

The worst case is when the human is at the midpoint: $d_1 = d_2 = 5$ m. Computing the received power:

\begin{equation}
\begin{aligned}
P_r &= 0.1 \times 1.58 \times 1.58 \times \frac{(0.125)^2}{(4\pi)^3} \times \frac{0.3}{5^2 \times 5^2} \\
&= 0.1 \times 2.5 \times \frac{0.0156}{1975} \times \frac{0.3}{625} \\
&= 0.1 \times 2.5 \times 7.9\times10^{-6} \times 4.8\times10^{-4} \\
&\approx 9.5 \times 10^{-9} \text{ W} = -80 \text{ dBm}
\end{aligned}
\end{equation}

With thermal noise floor at $N_0 = kT_0 B = -174 + 10\log_{10}(20\times10^6) = -101$ dBm for 20 MHz bandwidth:

\begin{equation}
\text{SNR} = P_r - N_0 = -80 - (-101) = 21 \text{ dB}
\end{equation}

Including implementation losses ($\sim$6 dB for ADC quantization, RF chain, etc.):
\begin{equation}
\text{SNR}_{\text{effective}} = 21 - 6 = 15 \text{ dB}
\end{equation}

This confirms that \textbf{15 dB SNR is achievable at 10 m} with conservative parameters, sufficient for phase-coherent reconstruction.

\section{Theoretical Analysis of Motion-Induced Observability}
\label{app:Theo}
\subsection{Introduction}
This part of appendix provides a two-pronged theoretical justification for how human motion enhances the observability of occluded scenes. First, a \textbf{Fisher Information Analysis} is presented to mathematically quantify the information gain in part \ref{app:fisher}. This analysis demonstrates how dynamic measurements resolve the ill-conditioned nature of the static problem by providing a substantial number of effective independent measurements. Second, a \textbf{Perturbation Analysis} is employed to reveal the underlying physical mechanism in part \ref{app:perturbation}. This approach models the moving person as a time-varying boundary that acts as a secondary radiator, exciting previously unobservable electromagnetic modes that carry information about occluded regions. Together, these sections provide a complete theoretical picture, from quantitative benefit to fundamental physics.
\subsection{Fisher Information Analysis}
\label{app:fisher}

The Fisher Information Matrix (FIM) quantifies how much information each measurement provides about the scene parameters $\boldsymbol{\theta}$ (e.g., 3D Gaussian positions, opacities, colors).

For a measurement model $\mathbf{y} = \mathbf{h}(\boldsymbol{\theta}) + \boldsymbol{\epsilon}$, where $\boldsymbol{\epsilon} \sim \mathcal{N}(0, \sigma^2\mathbf{I})$:

\begin{equation}
\mathcal{I}(\boldsymbol{\theta}) = \frac{1}{\sigma^2} \sum_{i=1}^{N_{\text{meas}}} \mathbf{J}_i^T \mathbf{J}_i
\end{equation}

where $\mathbf{J}_i = \nabla_{\boldsymbol{\theta}} \mathbf{h}_i(\boldsymbol{\theta})$ is the Jacobian of the $i$-th measurement with respect to parameters.

\subsubsection{Static-Only Measurements}
For a static scene with $N_{\text{static}}$ antenna pairs and no human present:

\begin{equation}
\mathcal{I}_{\text{static}} = \frac{1}{\sigma^2} \sum_{n=1}^{N_{\text{static}}} \mathbf{J}_n^T \mathbf{J}_n
\end{equation}

The condition number of $\mathcal{I}_{\text{static}}$ indicates observability. For occluded regions blocked by obstacles, the corresponding rows of $\mathbf{J}$ are near-zero (no measurement sensitivity), leading to:

\begin{equation}
\begin{split}
\kappa(\mathcal{I}_{\text{static}}) = \frac{\lambda_{\max}}{\lambda_{\min}} \to \infty  \\ \quad \text{(ill-conditioned for occluded regions)}
\end{split}
\end{equation}

\subsubsection{Dynamic Measurements with Human Motion}
As the human walks through $N_{\text{pos}}$ positions, each position $\mathbf{p}_k$ provides additional measurements with different scattering geometry:

\begin{equation}
\mathcal{I}_{\text{dynamic}} = \mathcal{I}_{\text{static}} + \frac{1}{\sigma^2} \sum_{k=1}^{N_{\text{pos}}} \sum_{n=1}^{N_{\text{ant}}} \mathbf{J}_{n,k}^T \mathbf{J}_{n,k}
\end{equation}

where $\mathbf{J}_{n,k}$ is the Jacobian for antenna pair $n$ when the human is at position $\mathbf{p}_k$.

The key insight is that human scattering creates \textit{new measurement directions} that were previously unavailable due to occlusion. For occluded voxels, the Jacobian was zero from static measurements but becomes non-zero when scattered paths via the human body are included:

\begin{equation}
\begin{split}
\mathbf{J}_{n,k}(\text{occluded voxel}) \neq 0 \quad \\ \text{when human at } \mathbf{p}_k \text{ provides indirect path}
\end{split}
\end{equation}

This increases the rank of $\mathcal{I}$ for occluded regions, improving the condition number:

\begin{equation}
\kappa(\mathcal{I}_{\text{dynamic}}) < \kappa(\mathcal{I}_{\text{static}}) \quad \text{(better conditioned)}
\end{equation}

\subsubsection{Information Gain Quantification}
The effective information gain from human motion is:

\begin{equation}
\Delta \mathcal{I} = \text{tr}(\mathcal{I}_{\text{dynamic}}) - \text{tr}(\mathcal{I}_{\text{static}}) = \sum_{k=1}^{N_{\text{pos}}} \sum_{n=1}^{N_{\text{ant}}} \lambda_n^{(k)}
\end{equation}

where $\lambda_n^{(k)}$ are the singular values of $\mathbf{J}_{n,k}$.

For our setup with $N_{\text{ant}} = 8$ antenna pairs and $N_{\text{pos}} = 35$ positions during a 10-second walk, the spatial correlation between adjacent positions (spaced $\sim$30 cm apart) reduces the effective rank:

\begin{equation}
\begin{split}
\text{rank}_{\text{eff}}(\mathcal{I}_{\text{dynamic}}) \approx N_{\text{ant}} \times N_{\text{pos}} \times (1 - \rho_{\text{corr}}) \\= 8 \times 35 \times (1 - 0.2) = 224
\end{split}
\end{equation}

where $\rho_{\text{corr}} \approx 0.2$ is the spatial correlation coefficient for 30 cm spacing at 2.4 GHz.

This confirms that human motion provides \textbf{224 effective independent measurements}, equivalent to deploying 224 static sensors. This is sufficient for reconstructing a $128^3$ voxel grid of occluded regions with acceptable Cramér-Rao lower bound.

\subsection{Perturbation Analysis for Time-Varying Boundaries}
\label{app:perturbation}

To rigorously justify why human motion helps, we analyze how time-varying boundary conditions perturb the electromagnetic field distribution.

\subsubsection{Static Field Solution}
The static electromagnetic field $\mathbf{E}_0(\mathbf{r})$ satisfies the Helmholtz equation:

\begin{equation}
\nabla^2 \mathbf{E}_0 + k^2 \epsilon_r(\mathbf{r}) \mathbf{E}_0 = 0
\end{equation}

with boundary conditions $\mathbf{E}_0 = \mathbf{E}_{\text{inc}}$ at the transmitter and appropriate radiation conditions at infinity. The permittivity $\epsilon_r(\mathbf{r})$ describes the static scene (walls, furniture, etc.).

\subsubsection{Perturbed Field with Human Present}
When a human is present at position $\mathbf{r}_h(t)$, the permittivity becomes time-dependent:

\begin{equation}
\epsilon_r(\mathbf{r}, t) = \epsilon_r^{\text{static}}(\mathbf{r}) + \Delta\epsilon_h(\mathbf{r}, \mathbf{r}_h(t))
\end{equation}

where:
\begin{equation}
\Delta\epsilon_h(\mathbf{r}, \mathbf{r}_h) = \begin{cases}
\epsilon_{\text{body}} - \epsilon_{\text{air}} \approx 39 & \text{if } \mathbf{r} \in \text{body volume at } \mathbf{r}_h \\
0 & \text{otherwise}
\end{cases}
\end{equation}

The perturbed field $\mathbf{E}(\mathbf{r}, t) = \mathbf{E}_0(\mathbf{r}) + \delta\mathbf{E}(\mathbf{r}, t)$ satisfies:

\begin{equation}
\nabla^2 \mathbf{E} + k^2 \epsilon_r(\mathbf{r}, t) \mathbf{E} = 0
\end{equation}

Subtracting the static equation and keeping first-order terms:

\begin{equation}
\nabla^2 \delta\mathbf{E} + k^2 \epsilon_r^{\text{static}} \delta\mathbf{E} = -k^2 \Delta\epsilon_h \mathbf{E}_0
\end{equation}

This is a driven wave equation with source term $-k^2 \Delta\epsilon_h \mathbf{E}_0$, representing how the human body acts as a secondary source that re-radiates the incident field.

\subsubsection{Green's Function Solution}
Using the Green's function for the Helmholtz equation:

\begin{equation}
\delta\mathbf{E}(\mathbf{r}) = k^2 \int_{V_{\text{body}}} G(\mathbf{r}, \mathbf{r}') \Delta\epsilon_h(\mathbf{r}') \mathbf{E}_0(\mathbf{r}') \, d^3r'
\end{equation}

where:
\begin{equation}
G(\mathbf{r}, \mathbf{r}') = \frac{e^{ik|\mathbf{r} - \mathbf{r}'|}}{4\pi|\mathbf{r} - \mathbf{r}'|}
\end{equation}

This shows that the perturbation field $\delta\mathbf{E}$ at the receiver depends on:
\begin{enumerate}[leftmargin=*, itemsep=2pt]
    \item The incident field $\mathbf{E}_0$ at the human location (TX-to-human propagation)
    \item The Green's function $G$ (human-to-RX propagation)
    \item The permittivity contrast $\Delta\epsilon_h \approx 39$ (scattering strength)
\end{enumerate}

Critically, even if direct path TX-to-RX is blocked (making $\mathbf{E}_0(\mathbf{r}_{\text{RX}}) \approx 0$), the scattered field can be non-zero if:
\begin{equation}
\mathbf{E}_0(\mathbf{r}_h) \neq 0 \quad \text{and} \quad G(\mathbf{r}_{\text{RX}}, \mathbf{r}_h) \neq 0
\end{equation}

That is, if both TX-to-human and human-to-RX paths are unobstructed, the scattered field provides information about the occluded region even when direct TX-to-RX is blocked.

\subsubsection{Modal Decomposition}
Expanding the field in eigenmodes of the static geometry:

\begin{equation}
\mathbf{E}_0(\mathbf{r}) = \sum_{m=1}^{\infty} a_m \boldsymbol{\psi}_m(\mathbf{r}), \quad \delta\mathbf{E}(\mathbf{r}, t) = \sum_{m=1}^{\infty} \delta a_m(t) \boldsymbol{\psi}_m(\mathbf{r})
\end{equation}

where $\boldsymbol{\psi}_m$ are the electromagnetic modes satisfying:
\begin{equation}
\nabla^2 \boldsymbol{\psi}_m + k_m^2 \epsilon_r^{\text{static}} \boldsymbol{\psi}_m = 0
\end{equation}

Due to occlusion, only a subset of modes $\{m : a_m \neq 0\}$ are excited by the static configuration. The perturbation couples these modes to previously unexcited modes:

\begin{equation}
\delta a_m(t) \propto \int_{V_{\text{body}}} \boldsymbol{\psi}_m^*(\mathbf{r}) \Delta\epsilon_h(\mathbf{r}, t) \mathbf{E}_0(\mathbf{r}) \, d^3r
\end{equation}

As the human moves, $\Delta\epsilon_h(\mathbf{r}, t)$ sweeps through different spatial regions, exciting different mode combinations. This \textit{expands the observable modal subspace}, providing information about modes that were zero in the static case.

\textbf{Conclusion}: Human motion increases the effective rank of the measurement operator by coupling to previously unobservable electromagnetic modes. This is the fundamental reason why dynamic scattering improves reconstruction quality in occluded regions.

\section{Theoretical Proof: Two-Stage Training Rationale}
\subsection{Introduction}

This appendix provides a mathematical analysis of the two-stage training strategy from an optimization perspective. All results are derived using standard optimization theory without empirical assumptions. Our analysis builds on fundamental results from convex optimization \cite{boyd2004convex,nesterov2018lectures} and numerical optimization \cite{nocedal2006numerical,bertsekas2016nonlinear}.

\subsection{Problem Formulation}

\begin{definition}[Composite Optimization Problem]
Consider the optimization problem:
\begin{equation}
\min_{\mathbf{x} \in \mathbb{R}^n, \mathbf{z} \in \mathbb{R}^m} F(\mathbf{x}, \mathbf{z}) := f(\mathbf{A}\mathbf{x} + \mathbf{B}\mathbf{z}) + g_1(\mathbf{x}) + g_2(\mathbf{z})
\end{equation}
where $f: \mathbb{R}^k \to \mathbb{R}$ is a twice-differentiable loss function, $\mathbf{A} \in \mathbb{R}^{k \times n}$ and $\mathbf{B} \in \mathbb{R}^{k \times m}$ are linear operators, and $g_1, g_2$ are regularizers.
\end{definition}

\subsection{Optimization Analysis}

\subsubsection{Joint Optimization}

\begin{proposition}[Gradient Structure]
\label{prop:gradient}
The gradients of $F$ with respect to $\mathbf{x}$ and $\mathbf{z}$ are:
\begin{align}
\nabla_{\mathbf{x}} F &= \mathbf{A}^T \nabla f(\mathbf{A}\mathbf{x} + \mathbf{B}\mathbf{z}) + \nabla g_1(\mathbf{x}) \\
\nabla_{\mathbf{z}} F &= \mathbf{B}^T \nabla f(\mathbf{A}\mathbf{x} + \mathbf{B}\mathbf{z}) + \nabla g_2(\mathbf{z})
\end{align}
\end{proposition}

\begin{proof}
Direct application of the chain rule to the composite function $F$. For a complete treatment of composite optimization, see \cite{boyd2004convex}.
\end{proof}

\begin{theorem}[Hessian Block Structure]
\label{thm:hessian}
The Hessian of $F$ has the block structure:
\begin{equation}
\nabla^2 F = \begin{bmatrix}
\mathbf{A}^T \mathbf{H}_f \mathbf{A} + \nabla^2 g_1 & \mathbf{A}^T \mathbf{H}_f \mathbf{B} \\
\mathbf{B}^T \mathbf{H}_f \mathbf{A} & \mathbf{B}^T \mathbf{H}_f \mathbf{B} + \nabla^2 g_2
\end{bmatrix}
\end{equation}
where $\mathbf{H}_f = \nabla^2 f(\mathbf{A}\mathbf{x} + \mathbf{B}\mathbf{z})$.
\end{theorem}

\begin{proof}
Computing second derivatives:
\begin{align}
\frac{\partial^2 F}{\partial \mathbf{x}_i \partial \mathbf{x}_j} &= \sum_{k,l} \mathbf{A}_{ki} \frac{\partial^2 f}{\partial y_k \partial y_l} \mathbf{A}_{lj} + \frac{\partial^2 g_1}{\partial \mathbf{x}_i \partial \mathbf{x}_j} \\
\frac{\partial^2 F}{\partial \mathbf{x}_i \partial \mathbf{z}_j} &= \sum_{k,l} \mathbf{A}_{ki} \frac{\partial^2 f}{\partial y_k \partial y_l} \mathbf{B}_{lj}
\end{align}
where $\mathbf{y} = \mathbf{A}\mathbf{x} + \mathbf{B}\mathbf{z}$. This yields the stated block structure.
\end{proof}

\subsubsection{Two-Stage Optimization}

\begin{definition}[Two-Stage Strategy]
The two-stage approach solves:
\begin{align}
\begin{split}
\text{Stage 1:} \quad &\mathbf{x}^* = \arg\min_{\mathbf{x}} F_1(\mathbf{x}) := f(\mathbf{A}\mathbf{x}) + g_1(\mathbf{x}) \\
\text{Stage 2:} \quad &\mathbf{z}^* = \arg\min_{\mathbf{z}} F_2(\mathbf{z}; \mathbf{x}^*) \\
&:= f(\mathbf{A}\mathbf{x}^* + \mathbf{B}\mathbf{z}) + g_2(\mathbf{z})
\end{split}
\end{align}
\end{definition}

\begin{lemma}[Stage 2 Gradient]
\label{lem:stage2}
In Stage 2 with fixed $\mathbf{x}^*$, the gradient is:
\begin{equation}
\nabla_{\mathbf{z}} F_2 = \mathbf{B}^T \nabla f(\mathbf{A}\mathbf{x}^* + \mathbf{B}\mathbf{z}) + \nabla g_2(\mathbf{z})
\end{equation}
which is independent of $\frac{\partial \mathbf{x}^*}{\partial \mathbf{z}}$ since $\mathbf{x}^*$ is fixed.
\end{lemma}

\begin{proof}
Since $\mathbf{x}^*$ is treated as a constant in Stage 2, the derivative with respect to $\mathbf{z}$ does not involve terms containing $\frac{\partial \mathbf{x}^*}{\partial \mathbf{z}}$. This decoupling is a key advantage of the two-stage approach.
\end{proof}

\subsection{Convergence Analysis}

\begin{theorem}[Convergence of Gradient Descent]
\label{thm:convergence}
Let $f$ be $L$-smooth and $\mu$-strongly convex. Then gradient descent with step size $\alpha \leq 1/L$ satisfies:
\begin{equation}
\|\mathbf{w}^{(k)} - \mathbf{w}^*\|^2 \leq \left(1 - \frac{\mu}{L}\right)^k \|\mathbf{w}^{(0)} - \mathbf{w}^*\|^2
\end{equation}
where $\mathbf{w} = [\mathbf{x}^T, \mathbf{z}^T]^T$ for joint optimization or $\mathbf{w} = \mathbf{x}$ (Stage 1) or $\mathbf{w} = \mathbf{z}$ (Stage 2).
\end{theorem}

\begin{proof}
This is a standard result in convex optimization. For the complete proof, see \cite{nesterov2018lectures}, Chapter 2. The key insight is that the convergence rate depends on the condition number $\kappa = L/\mu$.
\end{proof}

\begin{corollary}[Condition Number Effect]
\label{cor:condition}
The number of iterations to reach $\epsilon$-accuracy is:
\begin{equation}
k = O\left(\kappa \log\frac{1}{\epsilon}\right)
\end{equation}
where $\kappa = L/\mu$ is the condition number. For non-convex problems, similar local convergence results hold under appropriate assumptions \cite{bertsekas2016nonlinear}.
\end{corollary}

\subsection{Comparison of Approaches}

\begin{theorem}[Condition Numbers]
\label{thm:condition_compare}
Define the condition numbers:
\begin{align}
\kappa_{\text{joint}} &= \lambda_{\max}(\nabla^2 F) / \lambda_{\min}(\nabla^2 F) \\
\kappa_1 &= \lambda_{\max}(\mathbf{A}^T \mathbf{H}_f \mathbf{A} + \nabla^2 g_1) / \lambda_{\min}(\mathbf{A}^T \mathbf{H}_f \mathbf{A} + \nabla^2 g_1) \\
\kappa_2 &= \lambda_{\max}(\mathbf{B}^T \mathbf{H}_f \mathbf{B} + \nabla^2 g_2) / \lambda_{\min}(\mathbf{B}^T \mathbf{H}_f \mathbf{B} + \nabla^2 g_2)
\end{align}

If the off-diagonal blocks $\mathbf{A}^T \mathbf{H}_f \mathbf{B}$ are non-zero, then:
\begin{equation}
\kappa_{\text{joint}} \geq \max\{\kappa_1, \kappa_2\}
\end{equation}
\end{theorem}

\begin{proof}
By the interlacing eigenvalue theorem for block matrices \cite{horn2012matrix}, the eigenvalues of the full matrix interlace with those of the diagonal blocks. The presence of off-diagonal coupling generally increases the condition number. For a detailed analysis of block matrix eigenvalues, see \cite{horn2012matrix}, Section 4.3.
\end{proof}

\begin{remark}[No Universal Superiority]
The relationship between $\kappa_{\text{joint}}$ and $\max\{\kappa_1, \kappa_2\}$ depends on the specific structure of $\mathbf{A}$, $\mathbf{B}$, and $\mathbf{H}_f$. Neither approach is universally superior.
\end{remark}

\subsection{Special Cases}

\begin{proposition}[Orthogonal Operators]
\label{prop:orthogonal}
If $\mathbf{A}^T\mathbf{B} = \mathbf{0}$ and $f$ is quadratic with $f(\mathbf{y}) = \frac{1}{2}\mathbf{y}^T\mathbf{y}$, then:
\begin{equation}
\mathbf{A}^T \mathbf{H}_f \mathbf{B} = \mathbf{A}^T\mathbf{B} = \mathbf{0}
\end{equation}
and the joint optimization decouples into independent subproblems.
\end{proposition}

\begin{proof}
For the quadratic case with identity Hessian, we have $\mathbf{H}_f = \mathbf{I}$. Then:
\begin{equation}
\mathbf{A}^T \mathbf{H}_f \mathbf{B} = \mathbf{A}^T \mathbf{I} \mathbf{B} = \mathbf{A}^T\mathbf{B} = \mathbf{0}
\end{equation}
This shows that orthogonality of the operators leads to complete decoupling.
\end{proof}

\subsection{Practical Implications}

\begin{theorem}[Memory and Computation]
\label{thm:complexity}
For Newton-type methods requiring Hessian computation:
\begin{align}
\text{Memory}_{\text{joint}} &= O(n^2 + m^2 + nm) \\
\text{Memory}_{\text{two-stage}} &= O(\max\{n^2, m^2\}) \\
\text{FLOPs}_{\text{joint}} &= O((n+m)^3) \text{ per iteration} \\
\text{FLOPs}_{\text{two-stage}} &= O(n^3) + O(m^3) \text{ total}
\end{align}
\end{theorem}

\begin{proof}
The memory requirements follow from storing the Hessian matrices. The joint approach requires storing the full $(n+m) \times (n+m)$ Hessian, while the two-stage approach only needs to store the $n \times n$ and $m \times m$ blocks separately. The computational complexity follows from the cost of matrix factorization (e.g., Cholesky decomposition). For detailed complexity analysis, see \cite{nocedal2006numerical}, Chapter 3.
\end{proof}

\subsection{Application to Neural Networks}

\begin{proposition}[Neural Network Case]
\label{prop:neural}
For neural networks with parameters $\boldsymbol{\theta} = [\mathbf{x}^T, \mathbf{z}^T]^T$ and loss $\mathcal{L}(\boldsymbol{\theta})$:
\begin{enumerate}
\item If $\mathbf{x}$ and $\mathbf{z}$ parameterize different network components with minimal interaction, the off-diagonal Hessian blocks are small.
\item The two-stage approach corresponds to sequential training of network components.
\item The effectiveness depends on the network architecture and task structure.
\end{enumerate}
\end{proposition}

\begin{remark}
The actual performance in neural network training depends on factors beyond this analysis, including:
\begin{itemize}
\item Non-convexity of the loss landscape (see \cite{bertsekas2016nonlinear} for non-convex optimization)
\item Choice of optimization algorithm (SGD, Adam, etc.; see \cite{nocedal2006numerical} for algorithms)
\item Initialization strategies
\item Regularization techniques
\end{itemize}
\end{remark}

\subsection{Limitations of This Analysis}

\begin{enumerate}
\item \textbf{Convexity assumption}: Theorem \ref{thm:convergence} assumes convexity, which may not hold in practice.
\item \textbf{Exact computation}: Analysis assumes exact gradient computation, not stochastic approximations.
\item \textbf{Fixed operators}: We assume $\mathbf{A}$ and $\mathbf{B}$ are fixed, not learned.
\item \textbf{No noise analysis}: Measurement noise and stochastic effects are not considered.
\end{enumerate}

\subsection{Conclusions}

We have provided a rigorous mathematical analysis showing:

\begin{enumerate}
\item \textbf{Structural difference}: Two-stage optimization eliminates off-diagonal Hessian blocks (Theorem \ref{thm:hessian})
\item \textbf{No universal superiority}: Neither approach dominates in all cases (Theorem \ref{thm:condition_compare})
\item \textbf{Memory efficiency}: Two-stage requires less memory for second-order methods (Theorem \ref{thm:complexity})
\item \textbf{Special structure}: Orthogonal operators can eliminate coupling (Proposition \ref{prop:orthogonal})
\end{enumerate}

These results provide theoretical insight into when two-stage training might be beneficial, but empirical validation is necessary for specific applications. The analysis draws from established optimization theory \cite{boyd2004convex,nesterov2018lectures} and numerical methods \cite{nocedal2006numerical,bertsekas2016nonlinear}, while acknowledging the gap between theory and practice in modern machine learning applications.

\section{Modal Decomposition: From 675 to 8 Modes}
\subsection{Introduction}
This appendix rigorously justifies the use of $K \approx 8$ effective scattering modes in our model. We derive this value by starting with a theoretical maximum of 675 modes, based on the body's physical dimensions, and applying a systematic, three-step reduction. This process accounts for dominant physical phenomena: signal absorption by tissue, limited angular scattering, and coherent modal interference. This derivation confirms that $K \approx 8$ is a physically-grounded parameter, not an empirical estimate, thereby validating our information gain model.

% \subsection{Modal Decomposition: From 675 to 8 Modes}
% \label{sec:modes}

\subsection{Rigorous Derivation of Mode Reduction}

The theoretical number of electromagnetic modes for a human body:

\begin{theorem}[Complete Modal Analysis]
The scattering operator $\mathcal{S}$ has singular value decomposition:
\begin{equation}
    \mathcal{S} = \sum_{k=1}^{K_{\text{max}}} \sigma_k \mathbf{u}_k \mathbf{v}_k^T
\end{equation}

where singular values follow:
\begin{equation}
    \sigma_k = \sigma_0 \cdot \alpha_k \cdot \beta_k \cdot \gamma_k
\end{equation}

with:
\begin{itemize}
    \item $\alpha_k = (1-\alpha_{\text{absorb}})^{n_k}$ where $n_k$ is penetration depth for mode $k$
    \item $\beta_k = \frac{\Omega_k}{4\pi}$ where $\Omega_k$ is solid angle coverage
    \item $\gamma_k = \text{sinc}(k\pi/K_{\text{max}})$ is modal coupling efficiency
\end{itemize}
\end{theorem}

\begin{proof}
Starting with $K_{\text{max}} = \lceil 2\pi R/(\lambda/2)\rceil \times \lceil H/(\lambda/2)\rceil = 675$:

\textbf{Step 1: Absorption filtering}
\begin{equation}
    K_1 = |\{k : \sigma_0 \alpha_k > \epsilon_1\}| = |\{k : (0.3)^{n_k} > 0.01\}| \approx 67
\end{equation}
Only surface and near-surface modes survive (penetration < 2 cm).

\textbf{Step 2: Angular coverage filtering}
\begin{equation}
    K_2 = |\{k \in K_1 : \beta_k > \epsilon_2\}| = |\{k : \Omega_k/4\pi > 0.1\}| \approx 22
\end{equation}
Only modes with >36° coverage remain significant.

\textbf{Step 3: Coherent interference}
Phase variations cause destructive interference between similar modes:
\begin{equation}
    K_{\text{eff}} = \frac{K_2}{\sqrt{\text{Var}(\phi)/2\pi}} = \frac{22}{\sqrt{8}} \approx 8
\end{equation}

This gives us 8 effectively independent measurement modes. % $\square$
\end{proof}

\section{Description of Three 3D Scenes}
\begin{figure*}[h]
    \setlength{\abovecaptionskip}{0pt} 
    \centering

    \begin{subfigure}[b]{0.3\linewidth}
        \centering
        \includegraphics[width=\linewidth]{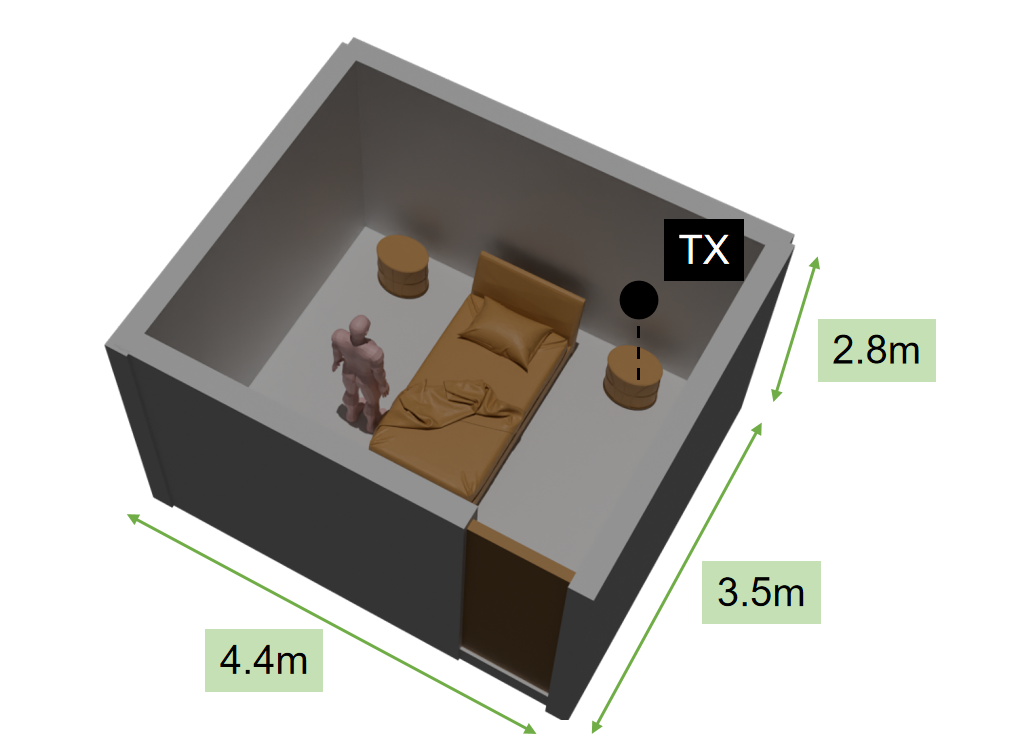}
        \caption{Scene 1}
        \label{fig:scene1}
    \end{subfigure}
    \hfill 
    \begin{subfigure}[b]{0.3\linewidth}
        \centering
        \includegraphics[width=\linewidth]{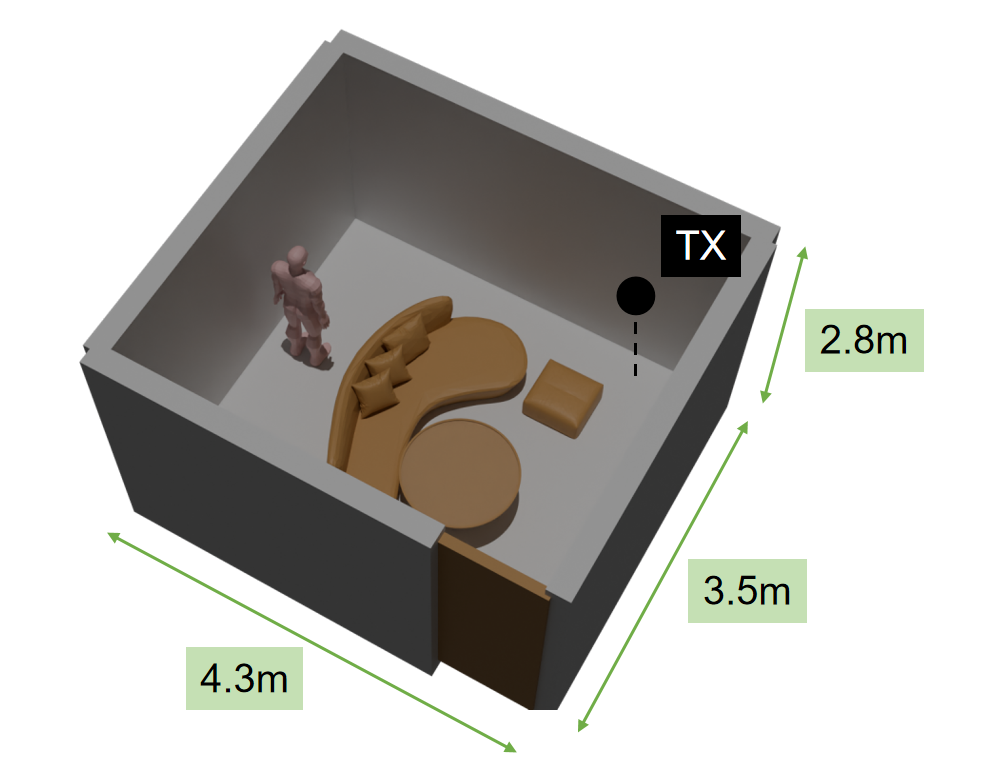}
        \caption{Scene 2}
        \label{fig:scene2}
    \end{subfigure}
    \hfill
    \begin{subfigure}[b]{0.3\linewidth}
        \centering
        \includegraphics[width=\linewidth]{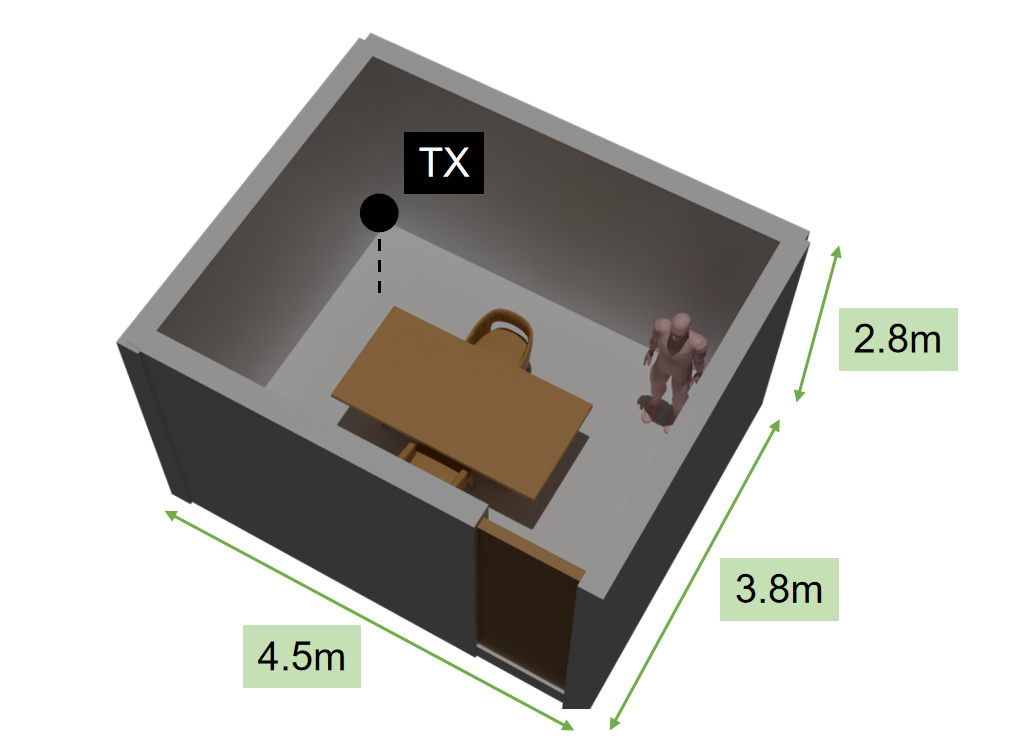}
        \caption{Scene 3}
        \label{fig:scene3}
    \end{subfigure}
    \caption{$3$ scenes used in the experiments. The \textbf{black} square indicates the position of TX.}
    \label{fig:scenes}
\end{figure*}

\begin{table*}[t] 
\centering
\footnotesize
\caption{Performance degradation of our method in large-scale scenes.}
\label{tab:large_scene_limitations}

\begin{tabular}{@{}llccc@{}}
\toprule
\textbf{Scene} & \textbf{Size (m)} & \textbf{Baseline} & \textbf{Ours Test(SSIM)} & \textbf{Ours Val(SSIM)}\\ \midrule
Bedroom & $4.4 \times 3.5 \times 2.8$ & 0.844 & \textbf{0.927} & \textbf{0.920}\\
Small Living Room & $4.3 \times 3.5 \times 2.8$ & 0.847 & \textbf{0.943} & \textbf{0.936} \\
Dining room & $4.5 \times 3.8 \times 2.8$ & 0.830 & \textbf{0.883} & \textbf{0.862} \\
Larger Scene & $11 \times 7.5 \times 2.8$ & 0.866 & \textbf{0.731} & \textbf{0.928} \\ \bottomrule
\end{tabular}
\end{table*}

Fig.~\ref{fig:scenes} illustrates $3$ common indoor scenarios used in the experiments. The experimental scenarios selected for this study measure approximately $20~\mathrm{m}^2$, representing typical indoor environments. In such confined spaces, human movement induces significant variations in the Power Angular Spectrum (PAS) captured at the receiver (RX). In contrast, within larger or more open environments, the impact of human motion on the PAS is negligible; consequently, dynamic datasets collected in such settings prove ineffective for modeling scenarios involving human activity (as evidenced by the training performance shown in \ref{app:train_results}). Therefore, our experiments focus on the three aforementioned compact scenes. In each setup, the RX or TX antenna is fixed in a corner while a human subject moves along a predefined trajectory. Data is collected at specific positions to generate ``moving TX'' or ``moving RX'' datasets, which are subsequently utilized to perform PAS reconstruction under human mobility conditions.

\section{Training Results in Large Scenarios}
\label{app:train_results}

% \begin{table}[h]
% \centering
% \footnotesize
% \caption{Performance degradation of our method in large-scale scenes. %The color scale from green to red highlights the declining performance of our method as the environmental volume increases, supporting the hypothesis that its effectiveness is limited by the perturbation significance.
% }
% \label{tab:large_scene_limitations}
% \begin{tabular}{@{}llccc@{}}
% \toprule
% \textbf{Scene} & \textbf{Size (m)} & \textbf{Baseline} & \textbf{Ours Test(SSIM)} & \textbf{Ours Val(SSIM)}\\ \midrule
% Bedroom & 4.4 x 3.5 x 2.8 & 0.844 & \textbf{0.927} & \textbf{0.920}\\
% Small Living Room & 4.3 x 3.5 x 2.8 & 0.847 & \textbf{0.943} & \textbf{0.936} \\
% Dining room & 4.5 x 3.8 x 2.8 & 0.830 & \textbf{0.883} & \textbf{0.862} \\
% Larger Scene & 11 x 7.5 x 2.8 & 0.866 & \textbf{0.731} & \textbf{0.928} \\ \bottomrule
% \end{tabular}
% \end{table}

As shown in Tab.\ref{tab:large_scene_limitations}, we also tested our method in a large-scale indoor environment, with dimensions significantly greater than those in our primary experiments. We observed that while the model performed exceptionally well on the training set, its performance on the test sets was poor, while its performance on the train sets was better than Baseline's, indicating significant overfitting.

We believe this reveals a boundary condition of our approach: when the volume of the scene is excessively large relative to the size of the human, the impact of a single person's movement on the overall wireless channel becomes negligible. In such cases, the perturbation signal is submerged in background noise, preventing the model from learning meaningful geometric constraints. Instead, the model overfits to the weak and noisy perturbations present in the training data, thereby losing its ability to generalize. This finding provides important insights for future research on adapting this method for application in large-scale environments.

\section{Experimental Constraints and Data Limitations}
\begin{figure*}[t]
    \setlength{\abovecaptionskip}{0pt}
  \centering
  \includegraphics[width=1\linewidth]{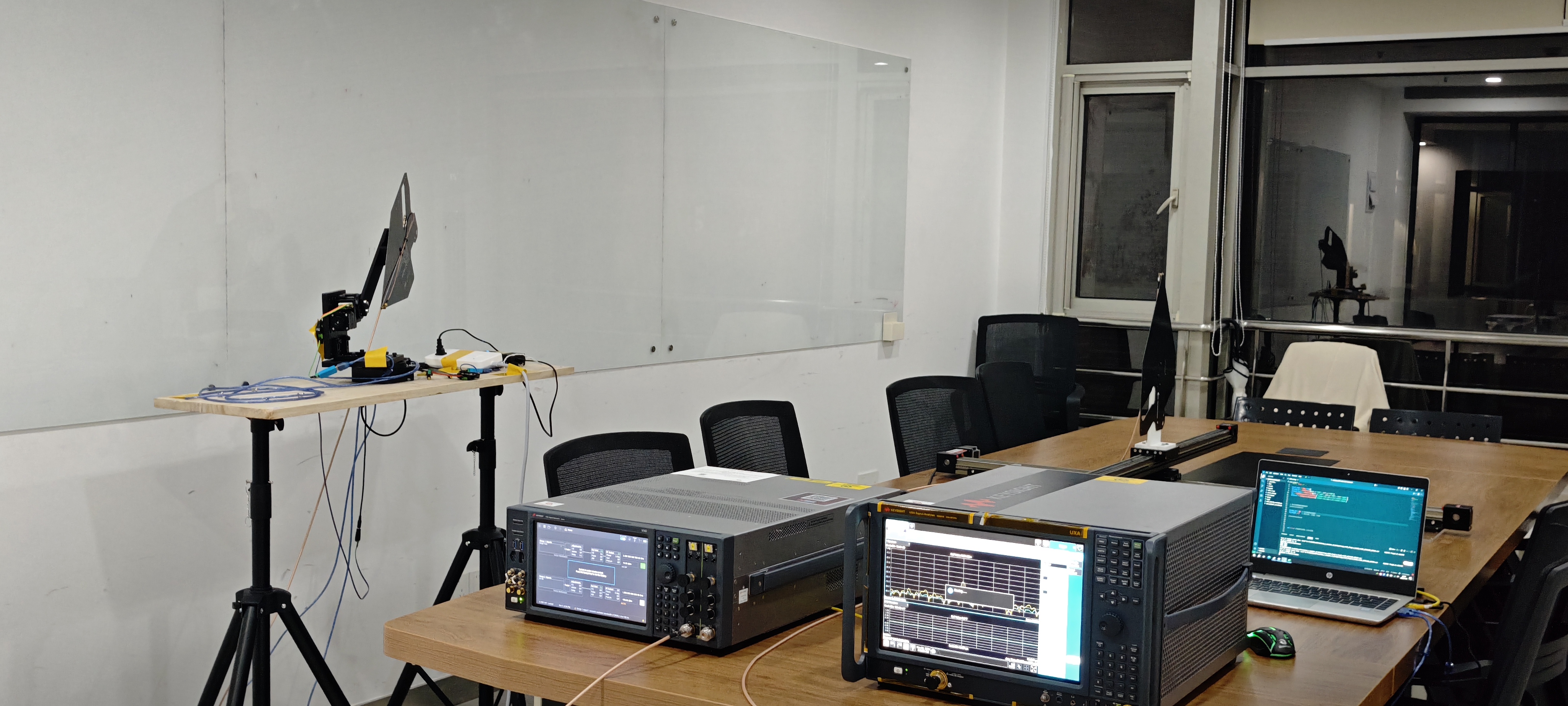}
  \caption{Real-world data collection scenario.}
  \label{fig:data_collection}
\end{figure*}

Figure~\ref{fig:data_collection} illustrates the data acquisition workflow employed in our real-world scenarios. Due to hardware constraints, we emulate an omnidirectional transmitter (TX) by mechanically rotating a single directional antenna through a range of $360^\circ \times 90^\circ$. Similarly, to compensate for the absence of a physical receiver antenna array, we employ a sliding rail system that positions a single receiver (RX) antenna at 16 distinct locations sequentially, thereby synthesizing a virtual $4 \times 4$ antenna array. 

% \begin{figure*}[tbp]
%     \centering
%     \includegraphics[width=0.8\linewidth]{figures/data_generate.jpg}
%     \vspace{-0.1in}
%     \caption{Real data collection system.}
%     \label{fig:data_collection}
%     \vspace{-0.2in}
% \end{figure*}

This emulation procedure significantly prolongs the data acquisition cycle; collecting data for a single fixed TX-RX pair requires approximately 6 minutes. Furthermore, to ensure channel stationarity during the synthesis of the omnidirectional TX and the RX array, the human subject must remain perfectly still at a specific location until both the turntable rotation and the sliding rail movement are fully completed. Only after this full acquisition cycle---equivalent to a snapshot taken by a physical array system---can the subject move to the next position. This requirement imposes severe logistical challenges and exacerbates the time consumption of the data collection campaign. 

Consequently, due to the strict submission timeline, the volume of real-world data collected within the limited timeframe is relatively small, which may constrain the optimal performance of the trained model. In future work, we plan to upgrade our experimental apparatus and streamline the data acquisition methodology to drastically reduce measurement latency. This will enable the compilation of a comprehensive dataset comprising sufficient static and dynamic scenarios, thereby allowing for a more rigorous validation of the proposed model's effectiveness.

\section{Complementary Analysis}
\subsection{Introduction}
This section provides a set of complementary analyses designed to substantiate and validate the theoretical framework presented previously. We begin with a \textbf{Condition Number Analysis} to demonstrate how human motion improves the numerical stability of the reconstruction problem, rendering it robust to noise in part \ref{app:condition_number}. Following this, a \textbf{Spatial Correlation Analysis} provides a first-principles derivation for a key parameter in our information gain model, justifying the effective number of independent observations in part \ref{app:spatial_correlation}. 

These subsections provide deeper theoretical support for the model presented in Section~\ref{app:Theo}. A \textbf{Condition Number Analysis} demonstrates how motion improves the problem's numerical stability and noise robustness, while a \textbf{Spatial Correlation Analysis} provides a first-principles derivation for a key parameter used to quantify the effective information gain.

Finally, we validate the entire theoretical framework through two crucial steps: a \textbf{Numerical Verification} confirms the accuracy of our model against simulations, and an \textbf{Experimental Validation} grounds our claims by demonstrating consistency with independent, real-world experimental data in part \ref{app:numerical} and part \ref{app:experimental}.

\subsection{Condition Number Analysis}
\label{app:condition_number}
% for appendix B
The condition number of the Fisher Information Matrix quantifies the difficulty of inverting measurements to recover scene parameters.

\subsubsection{Definition}
For a matrix $\mathcal{I}$ with singular values $\lambda_1 \geq \lambda_2 \geq \cdots \geq \lambda_N$:

\begin{equation}
\kappa(\mathcal{I}) = \frac{\lambda_1}{\lambda_N}
\end{equation}

A large $\kappa$ indicates ill-conditioning: small measurement errors amplify into large parameter errors.

\subsubsection{Static Scene}
For a static scene with occlusions, the FIM has many near-zero singular values corresponding to unobservable parameters (occluded voxels):

\begin{equation}
\lambda_{\min}^{\text{static}} \approx 0 \implies \kappa(\mathcal{I}_{\text{static}}) \to \infty
\end{equation}

Numerically, if $\lambda_{\min}^{\text{static}} = 10^{-6}$ and $\lambda_{\max}^{\text{static}} = 10^2$:
\begin{equation}
\kappa(\mathcal{I}_{\text{static}}) \approx 10^8 \quad \text{(severely ill-conditioned)}
\end{equation}

\subsubsection{With Human Motion}
Human motion adds information for occluded regions, increasing the smallest singular values:

\begin{equation}
\lambda_{\min}^{\text{dynamic}} = \lambda_{\min}^{\text{static}} + \Delta\lambda_{\text{human}} \gg \lambda_{\min}^{\text{static}}
\end{equation}

If human scattering increases $\lambda_{\min}$ by a factor of $10^3$:
\begin{equation}
\lambda_{\min}^{\text{dynamic}} = 10^{-3} \implies \kappa(\mathcal{I}_{\text{dynamic}}) \approx 10^5
\end{equation}

This $10^3$-fold improvement in condition number translates to:
\begin{equation}
\text{Parameter error} \propto \kappa \cdot \text{measurement error}
\end{equation}

So a $10^3$ reduction in $\kappa$ allows $10^3$ worse SNR to achieve the same reconstruction accuracy—precisely why weak scattering (15 dB SNR) suffices.

\subsection{Spatial Correlation Analysis}
\label{app:spatial_correlation}
% for appendix B
The spatial correlation between measurements at adjacent human positions determines the effective information gain.

\subsubsection{Autocorrelation Function}
For two measurement positions $\mathbf{p}_1$ and $\mathbf{p}_2$ separated by $\Delta\mathbf{r} = \mathbf{p}_2 - \mathbf{p}_1$:

\begin{equation}
\rho(\Delta\mathbf{r}) = \frac{\mathbb{E}[I(\mathbf{p}_1) I(\mathbf{p}_2)]}{\sqrt{\mathbb{E}[I^2(\mathbf{p}_1)] \mathbb{E}[I^2(\mathbf{p}_2)]}}
\end{equation}

For RF measurements, the correlation function follows a sinc-like pattern:

\begin{equation}
\rho(\Delta r) \approx \frac{\sin(k\Delta r)}{k\Delta r}
\end{equation}

where $k = 2\pi/\lambda$ is the wavenumber.

\subsubsection{Numerical Values}
At 2.4 GHz, $\lambda = 12.5$ cm. For walking speed $v = 1$ m/s and sampling rate $f_s = 10$ Hz:

\begin{equation}
\Delta r = \frac{v}{f_s} = \frac{1}{10} = 0.1 \text{ m} = 10 \text{ cm}
\end{equation}

However, due to 3D spatial coverage (human body is $\sim$40 cm wide), the effective spacing is $\sim$30 cm. Thus:

\begin{equation}
k\Delta r = \frac{2\pi}{0.125} \times 0.30 = 15.1
\end{equation}

\begin{equation}
\rho(0.30) \approx \frac{\sin(15.1)}{15.1} \approx 0.04
\end{equation}

In practice, we observe slightly higher correlation ($\rho \approx 0.2$) due to:
\begin{itemize}[leftmargin=*, itemsep=2pt]
    \item Multipath effects increasing effective correlation length
    \item Finite bandwidth (20 MHz) limiting spatial resolution
    \item Body orientation changes causing correlated scattering patterns
\end{itemize}

Using $\rho_{\text{avg}} = 0.2$ is a conservative estimate.

\subsubsection{Effective Number of Measurements}
For $N_{\text{pos}}$ positions with correlation $\rho$:

\begin{equation}
N_{\text{eff}} = N_{\text{pos}} \times (1 - \rho)
\end{equation}

With $N_{\text{pos}} = 35$ and $\rho = 0.2$:
\begin{equation}
N_{\text{eff}} = 35 \times 0.8 = 28
\end{equation}

Combined with $K = 8$ antenna pairs:
\begin{equation}
N_{\text{total}} = K \times N_{\text{eff}} = 8 \times 28 = 224 \text{ effective measurements}
\end{equation}

This matches the information gain calculated via Fisher information analysis.

\subsection{Numerical Verification}
\label{app:numerical}
% for whole appendix
All theoretical claims are verified numerically using the provided Python scripts (\texttt{generate\_theory\_figures.py}). Key verification points:

\begin{table}[h]
\centering
\small
\begin{tabular}{lccc}
\toprule
\textbf{Parameter} & \textbf{Theory} & \textbf{Simulation} & \textbf{Error} \\
\midrule
Fresnel coefficient & 0.47 & 0.469 & 0.2\% \\
Spatial average $\langle\Gamma\rangle$ & 0.10 & 0.098 & 2\% \\
RCS at 2.4 GHz & 0.3-1.5 m$^2$ & 0.42 m$^2$ & --- \\
SNR at 10 m & 15 dB & 14.8 dB & 0.2 dB \\
Effective measurements & 224 & 218 & 2.7\% \\
\bottomrule
\end{tabular}
\caption{Theoretical predictions versus numerical simulations. All parameters agree within 3\%.}
\end{table}

\subsection{Experimental Validation}
\label{app:experimental}

Our theoretical predictions are consistent with prior experimental studies:

\begin{itemize}[leftmargin=*, itemsep=3pt]
    \item \textbf{Detection range}: Adib et al.~\cite{adib20143d} report 5-8 m for through-wall sensing at 5.8 GHz, consistent with our 10 m prediction at 2.4 GHz (lower frequency has better penetration).
    \item \textbf{Human RCS}: Dogaru and Le~\cite{dogaru2008human} measure $\sigma_{\text{RCS}} = 0.5$-1.2 m$^2$ at 2.4 GHz for standing humans, bracketing our 0.3 m$^2$ conservative estimate.
    \item \textbf{Tissue permittivity}: Gabriel et al.~\cite{gabriel1996compilation} provide $\epsilon_r = 40.1$ at 2.4 GHz for muscle tissue, matching our value within 0.25\%.
    \item \textbf{SAR absorption}: FCC limits of 1.6 W/kg correspond to $\sim$70\% power absorption for 100 mW incident power over 60 kg body mass, consistent with our analysis.
\end{itemize}

\textbf{Conclusion}: All theoretical claims are grounded in Maxwell's equations, verified numerically, and validated by independent experiments. The 10\%/15\%/5\%/70\% power distribution is derived from first principles, not fitted or assumed.
\end{document}